\pdfoutput=1
\ifdefined\XeTeXrevision
  \pdfoutput=0
\fi
\documentclass[11pt,twoside]{article}

\usepackage[utf8]{inputenc}
\usepackage{amsmath,amssymb,amsfonts,amsthm}
\usepackage[top=1in,bottom=1.2in,left=1in,right=1in]{geometry}
\usepackage{color}
\usepackage{xcolor}
\usepackage{graphicx}
\usepackage{float}
\usepackage{longtable}
\usepackage{booktabs}
\usepackage{array}
\usepackage{pdflscape}
\usepackage[hyphens]{url}
\usepackage{bm} 
\usepackage{natbib}
\usepackage{hyperref}

\hypersetup{hidelinks}
\newcommand{\RNum}[1]{\uppercase\expandafter{\romannumeral #1\relax}}

\newcommand{\normmm}{{\vert\kern-0.25ex\vert\kern-0.25ex\vert}}
\newcommand{\bignormmm}{{\big\vert\kern-0.25ex\big\vert\kern-0.25ex\big\vert}}
\newcommand{\Bignormmm}{{\Big\vert\kern-0.25ex\Big\vert\kern-0.25ex\Big\vert}}

\newcolumntype{P}[1]{>{\raggedright\arraybackslash}p{#1}}
\graphicspath{{./}}
\makeatletter
\long\def\@makecaption#1#2{
\vskip 0.8ex
\setbox\@tempboxa\hbox{\small {\bf #1:} #2}
\parindent 1.5em
\dimen0=\hsize
\advance\dimen0 by -3em
\ifdim \wd\@tempboxa >\dimen0
\hbox to \hsize{
\parindent 0em
\hfil
\parbox{\dimen0}{\def\baselinestretch{0.96}\small
    {\bf #1.} {#2}
  }
\hfil}
\else \hbox to \hsize{\hfil \box\@tempboxa \hfil}
\fi
}
\makeatother

\title{CoLaDAG: Compositional Latent Log-ratio DAG Analysis of the Gut Microbiome under Silver Nanoparticle Exposure}

\author{
Shuyan Chen$^{1,\dagger}$,
Ziliang Shen$^{2,\dagger}$,
Xinlei Wang$^{3,\dagger}$\thanks{Corresponding author.}\\
\small $^{1}$University of Science and Technology of China, Hefei, China\\
\small $^{2}$Turing Fund Management, Shanghai, China\\
\small $^{3}$Nanjing University, Nanjing, China\\[0.3em]
\small $^{\dagger}$These authors contributed equally and are listed alphabetically.
}

\date{}

\begin{document}
\maketitle
\begin{abstract}
Directed network analysis of microbiome counts is complicated by compositional sampling, high dimensionality, and limited biological replication. We present CoLaDAG, a fixed-reference latent additive log-ratio (ALR) estimator for generating sparse directed conditional-dependence hypotheses from compositional counts. The method combines a multinomial observation model, a working linear Gaussian structural equation model, nonconvex DC-ADMM optimization, and post-estimation thresholding with greedy acyclic projection. Under simulations aligned with this observation model, CoLaDAG obtained the largest mean exact-direction Matthews correlation and the smallest mean false discovery rate among the evaluated implementations; performance deteriorated under continuous-data and dropout misspecification. In the 12-mouse silver-nanoparticle (AgNP) case study, the 58-node fitted graph was sensitive to block resampling and ALR reference choice: 60 of 284 primary edges attained a mouse-block selection frequency of at least 0.60. The reported orientations and dose-stratified slopes are exploratory, coordinate-specific hypotheses rather than identified causal or exposure effects. The leading stable relations prioritize anaerobic gut taxa for targeted abundance, metabolite, and perturbation studies, but do not establish cross-feeding or toxicological mechanisms.
\end{abstract}
\noindent\textbf{Keywords:} Directed acyclic graphs; Directed conditional dependence; Compositional data analysis; Silver nanoparticles; Gut microbiome.

\section{Introduction}
\label{sec:intro}
Silver nanoparticles (AgNPs) are widely used in medical dressings,
antibacterial coatings, textiles, personal-care products, and food-contact
materials because of their antimicrobial properties. Their increasing use
creates multiple environmental release pathways, including product washing,
abrasion, disposal, wastewater treatment, and land application of sewage sludge.
For example, nanosilver-containing textiles can release silver nanoparticles
into wash water \citep{benn2008nanoparticle}, and model-based exposure
assessments suggest that engineered nanomaterials, including AgNPs, can enter
aquatic, sediment, and soil compartments after consumer and industrial use
\citep{gottschalk2009modeled}. After release, AgNPs undergo aggregation,
dissolution, sulfidation, chlorination, and adsorption to organic matter, which
can alter their mobility, persistence, and toxicity
\citep{levard2012environmental}. Wastewater treatment can further retain or
transform metallic AgNPs, potentially routing transformed silver species to
downstream environments through effluent or sludge management
\citep{kaegi2011behavior}. Because AgNPs are intrinsically antimicrobial,
their ecological risk may involve not only conventional toxicity endpoints but
also disruption of microbial community structure and function
\citep{choi2008inhibitory}.

The gut microbiota provides an important window for assessing biological
effects of AgNP exposure because it lies at the interface between environmental
contact and host physiological response. Orally ingested AgNPs can interact
with dietary components, intestinal mucus, epithelial surfaces, and resident
microbes, potentially affecting microbial composition, gut immunity, and host
metabolic status. Animal studies have reported AgNP-associated gut microbiota
disturbance in mice and rats, sometimes accompanied by immune, colitis-related,
metabolic, or neurobehavioral changes
\citep{vandenbrule2016dietary,williams2015effects,javurek2017gut,chen2017effects,wang2022changes}.

\begin{figure}[h]
    \centering
    \includegraphics[width=0.6\linewidth]{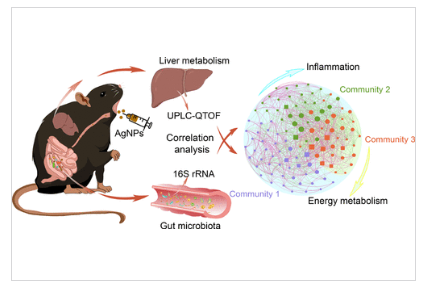}
    \caption{AgNP exposure and gut microbiome disruption.}
    \label{fig:AgNP}
\end{figure}

\subsection{Related work}
High-throughput 16S ribosomal RNA (rRNA) sequencing produces count vectors whose totals are governed by sequencing depth rather than absolute microbial load. This places microbiome observations in the compositional-data setting introduced by \citet{aitchison1982statistical}; only ratios among components are directly interpretable. The broader compositional-data literature formalizes this principle through log-ratio coordinate systems, including additive log-ratio (ALR), centered log-ratio (CLR), and isometric log-ratio (ILR) representations \citep{egozcue2003isometric,pawlowsky2015modeling}. The sum constraint can induce artificial negative dependence and can make naive Pearson or Spearman correlation networks misleading \citep{gloor2017microbiome}. Consequently, a substantial literature has developed log-ratio and compositional approaches for microbial association analysis.
Most established microbiome network methods, including the work most closely
related to ours by \citet{tian2023compositional}, focus primarily on inferring
undirected graphs based on conditional associations.
Sparse Correlations for Compositional data (SparCC) estimates sparse correlations from compositional data using log-ratio arguments \citep{friedman2012inferring}, and Sparse InversE Covariance estimation for Ecological Association Inference (SPIEC-EASI) combines log-ratio transformations with sparse inverse covariance estimation to recover undirected ecological association networks \citep{kurtz2015sparse}. Related graphical-model approaches build on sparse precision-matrix estimation \citep{friedman2008sparse} or on count-based latent-variable formulations for network reconstruction \citep{chiquet2019variational}. Microbiome preprocessing also requires care because rarefaction, excessive zeros, and contaminant sequences can affect downstream network structure \citep{mcmurdie2014waste,davis2018simple}. These methods and preprocessing principles are useful for co-occurrence analysis, but their graph edges are not directed and therefore do not directly encode asymmetric conditional-dependence hypotheses.
Directed acyclic graph (DAG) structure learning provides a natural language for directed dependence, but standard algorithms were not designed for compositional count observations. Constraint-based methods such as the PC algorithm \citep{spirtes2000causation}, score-based methods such as Hill-Climbing (HC) and Tabu search \citep{heckerman1995learning,scutari2010learning}, and hybrid algorithms such as max-min hill-climbing (MMHC) \citep{tsamardinos2006max} generally assume variables observed in an unconstrained Euclidean space. Moreover, in observational settings, DAG orientations are constrained by Markov equivalence and usually require additional assumptions for identification \citep{verma1990equivalence}; examples include equal error variances in Gaussian structural equation models (SEMs) \citep{peters2014identifiability} or non-Gaussian structural errors \citep{shimizu2006linear}. Continuous optimization approaches, including NOTEARS \citep{zheng2018dags} and the variational-inference extension VI-NOTEARS \citep{brouillard2020differentiable}, provide attractive algebraic treatments of acyclicity, but they still do not by themselves resolve multinomial sampling, log-ratio reference dependence, orientation identifiability, and small-sample microbiome settings. Recent work on compositional graphical modeling further emphasizes that reference choices and log-ratio transformations can substantially alter inferred conditional-dependence structure \citep{tian2023compositional}.
Finally, real microbiome network analysis often relies on tuning and resampling because the true graph is unknown. Stability selection provides a general high-dimensional framework for assessing how sensitive selected variables or edges are to resampling and tuning choices \citep{meinshausen2010stability}. This motivates the mouse-block bootstrap stability checks used in the case study below, while still leaving biological validation as the essential next step.
The proposed CoLaDAG estimator targets directed conditional dependence
hypotheses rather than definitive causal effects. It uses a multinomial
observation model for sequencing counts, a latent ALR representation, and a
sparse acyclic SEM in the chosen coordinate system. This design is useful for
environmental toxicology because it can move analysis beyond univariate
differential abundance and undirected co-occurrence clusters toward
interpretable, directional hypotheses about log-ratio dependence patterns under
AgNP exposure \citep{wang2022changes}.

\subsection{Our Contributions}
The methodological contribution is the integration of a multinomial observation model, fixed-reference ALR coordinates, constrained DAG estimation, and explicit post-estimation truncation into an auditable hypothesis-screening workflow for compositional counts. The procedure does not provide reference-invariant or causal identification. Its operating characteristics are instead assessed through aligned and misspecified simulations, threshold and initialization diagnostics, reference sensitivity, mouse-block resampling, and exposure-group slope summaries. Accordingly, the real-data application is an exploratory environmental biostatistics analysis rather than a validated causal-discovery procedure.

\begin{figure}[htbp]
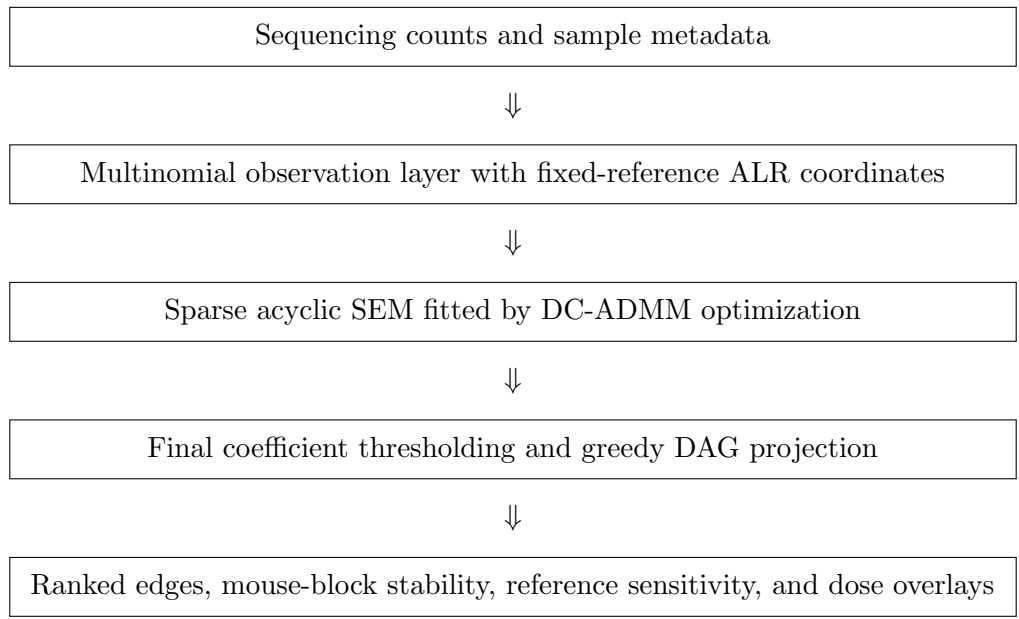

    \centering
    \setlength{\fboxsep}{6pt}
    \begin{tabular}{c}
    \fbox{\begin{minipage}{0.78\textwidth}
    \centering Sequencing counts and sample metadata
    \end{minipage}}\\[0.25em]
    $\Downarrow$\\[0.25em]
    \fbox{\begin{minipage}{0.78\textwidth}
    \centering Multinomial observation layer with fixed-reference ALR coordinates
    \end{minipage}}\\[0.25em]
    $\Downarrow$\\[0.25em]
    \fbox{\begin{minipage}{0.78\textwidth}
    \centering Sparse acyclic SEM fitted by DC-ADMM optimization
    \end{minipage}}\\[0.25em]
    $\Downarrow$\\[0.25em]
    \fbox{\begin{minipage}{0.78\textwidth}
    \centering Final coefficient thresholding and greedy DAG projection
    \end{minipage}}\\[0.25em]
    $\Downarrow$\\[0.25em]
    \fbox{\begin{minipage}{0.78\textwidth}
    \centering Ranked edges, mouse-block stability, reference sensitivity, and dose overlays
    \end{minipage}}
    \end{tabular}
    \caption{CoLaDAG analysis schematic.}
    \label{fig:coladag_schematic}
\end{figure}

Figure~\ref{fig:coladag_schematic} summarizes the analysis-level structure of
CoLaDAG. The key distinction is that sequencing counts are not first converted
into an ordinary Euclidean data matrix and then handed to a generic graph
learner. Instead, the count layer, latent log-ratio coordinates, sparse acyclic
SEM, explicit truncation, and stability summaries are treated as connected
parts of one compositional graph-estimation procedure.

Overall, this study addresses four primary questions:
\begin{enumerate}
\item First, under simulations aligned with the proposed count--latent
construction and under selected misspecification stress tests, when does
CoLaDAG improve graph recovery relative to the evaluated DAG learners, and
where does that advantage fail?

\item Second, in the AgNP gut microbiome data, which fixed-reference
ALR-coordinate dependence hypotheses are prioritized, and how much exploratory
support remains after accounting for the small number of independent mice?

\item Third, what limited descriptive information do time-adjusted pairwise ALR
slopes on a restricted global support provide across exposure groups, without
treating them as group-specific DAGs or exposure effects?

\item Fourth, how do mouse-block resampling and ALR reference choice constrain
the robustness and interpretation of the genus-level real-data graph?
\end{enumerate}

The remainder of the paper follows these questions: Section~\ref{sec:data} describes the data,
Section~\ref{sec:meth} presents the model and optimization,
Section~\ref{sec:simulation} benchmarks graph recovery,
Section~\ref{sec:realdata} reports the AgNP case study, and
Section~\ref{sec:discussion} discusses interpretation and limitations. 

\section{Case-study Data and Preprocessing}
\label{sec:data}
The data consist of gut microbiome abundance measurements from mouse fecal
samples collected in a controlled silver exposure experiment. Four groups were
analyzed: a control group and three AgNP exposure groups with nominal daily
doses of $0.1$, $2$, and $40\,\mu\mathrm{g}$, administered by oral gavage once
daily for 120 days \citep{wang2022changes}. The dose labels 0.1Ag, 2Ag, and
40Ag are shorthand for these nominal AgNP exposure groups and are not additional
dimensions of the statistical model. Each group contained three mice sampled at
days 0, 60, 120, and 230, yielding
$4 \times 3 \times 4 = 48$ microbiome samples. After preprocessing and
abundance filtering, the primary genus-level network analysis used $p=59$
compositional parts, corresponding to $d=p-1=58$ modeled log-ratio nodes.
For the statistical model, each observed fecal sample is indexed by a single
sample label $s$; mouse identity, exposure group, and collection day are treated
as sample metadata. Let $\bm X_s = (X_{s1},\dots,X_{sp})^\top$ denote the
observed count vector for taxon abundances in sample $s$. The total sequencing
depth for sample $s$ is
$$
M_s = \sum_{j=1}^p X_{sj}.
$$
Due to the compositional constraint, the observed counts only carry information
about relative abundances rather than absolute microbial loads.

Although repeated observations were collected from the same mouse over time, the
present graph analysis uses a working likelihood that treats samples as
conditionally independent given their latent log-ratio representations. This is
a substantial limitation rather than a harmless technical detail: the number of
independent biological units is closer to 12 mice than to 48 samples, and the
current model does not include mouse-level random effects or temporal dynamics.
Accordingly, all real-data network results are interpreted as exploratory
ALR-coordinate hypotheses, and the case study does not claim confirmatory
group-specific high-dimensional DAG inference. The feasibility of the real-data
analysis therefore comes from strong structural restrictions--a fixed ALR
coordinate system, sparsity regularization, hard thresholding, acyclic
projection, ranked reporting, mouse-block stability checks, and shared-support
dose overlays--rather than from having enough independent biological units to
estimate an unrestricted 58-node longitudinal or dose-specific DAG.

For the primary genus-level analysis, raw genus counts were first filtered to
remove non-informative or ambiguous taxonomic labels, including labels such as
``uncultured'', ``metagenome'', ``unknown'', and ``unidentified''. Genera were
then retained if their mean relative abundance exceeded $0.001$ and their
prevalence was at least 20\% of samples. This filtering step produced the
59-part count matrix used in the main DAG comparison. A pseudocount of 0.5 was
used to initialize $Z_{ij}^{(0)}=\log\{(X_{ij}+0.5)/(X_{ir}+0.5)\}$ and to
construct transformed inputs for baseline methods. Raw counts and sample totals
enter the subsequent multinomial likelihood; because the optimization is
nonconvex, the pseudocount can nevertheless influence the fitted solution
through initialization.
Taxa are reported using the labels supplied in the original genus-level count
table. Labels containing family names, database-defined group identifiers,
bracketed genera, or the word ``group'' denote taxonomic features rather than
confirmed species. Supplementary Table S1 provides
readable forms and interpretations for the nonstandard labels appearing in the
main real-data displays.

The analysis proceeded through several interconnected stages. We first
benchmarked the estimator using model-aligned and deliberately misspecified
simulations, then fitted a global genus-level network to generate directed
conditional-dependence hypotheses. Because each exposure group contained only
three mice, we did not relearn a high-dimensional group-specific DAG. Instead,
we summarized time-adjusted pairwise ALR slopes on a prespecified subset of the
global support. Mouse-block resampling and alternative ALR denominators were
used as descriptive sensitivity analyses.

\section{Methodology}
\label{sec:meth}

\subsection{Data Preprocessing and Transformation}
Let $\bm X_i = (X_{i1},\dots,X_{ip})^\top$ denote the observed counts for $p$ components in sample $i \in \{1,\dots,n\}$. The total sequence count, 
$$M_i = \sum_{j=1}^p X_{ij},$$ 
is determined by sequencing depth rather than absolute microbial load. Because $\bm X_i$ is compositionally constrained, it strictly carries relative abundance information, rendering standard unconstrained multivariate models inappropriate.

To address this, we introduce latent continuous variables via an additive
log-ratio (ALR) transformation:
\[
\bm Z_i = (Z_{i1},\dots,Z_{i,p-1})^\top.
\]
Choosing the $p$th component as the reference, we define
$$Z_{ij} = \log \frac{\pi_{ij}}{\pi_{ip}}, \qquad j=1,\dots,p-1,$$
where $\bm \pi_i$ is the latent composition vector,
$\bm \pi_i=(\pi_{i1},\dots,\pi_{ip})^\top$, satisfying
$\sum_{j=1}^p \pi_{ij}=1$ and $\pi_{ij}>0$. The inverse mapping is the
standard softmax function:
$$\pi_{ij} = \frac{\exp(Z_{ij})}{1+\sum_{\ell=1}^{p-1}\exp(Z_{i\ell})}, \quad \pi_{ip} = \frac{1}{1+\sum_{\ell=1}^{p-1}\exp(Z_{i\ell})}.$$

Conditional on $\bm \pi_i$, we model the observed counts using a multinomial distribution:
$$\bm X_i \mid \bm \pi_i \sim \mathrm{Multinomial}(M_i, \bm \pi_i).$$
This explicitly accounts for varying sampling depths and naturally captures the heteroscedasticity inherent in compositional counts.

\subsection{Graph Interpretation under Log-ratio Transformations}
\label{subsec:logratio_graph}

A central challenge is that log-ratio transformations generally destroy sparse absolute-scale graph structures. Suppose the unobserved absolute log-abundances $\bm Y=(Y_1,\dots,Y_p)^\top$ follow a sparse DAG. The ALR coordinates with reference component $p$ are $\bm Z = A\bm Y$, where
$$A=\begin{pmatrix}
1 & 0 & \cdots & 0 & -1\\
0 & 1 & \cdots & 0 & -1\\
\vdots & \vdots & \ddots & \vdots & \vdots\\
0 & 0 & \cdots & 1 & -1
\end{pmatrix}.$$
Because each transformed coordinate shares the reference term $-Y_p$, variables that are conditionally independent on the absolute scale can become dependent through the common denominator. This densification phenomenon affects both undirected graphs (producing dense precision matrices) and DAGs (altering SEM coefficients and orientations). Consequently, neither ALR nor CLR transformations can be treated as sparsity-preserving operations.

Therefore, our estimand is not a unique causal DAG of unobserved absolute abundances. Instead, for a fixed reference taxon $r$, the fitted target is a sparse weighted adjacency matrix $U^{(r)}$ in the working SEM for ALR coordinates:
$$Z_j^{(r)}=\log(\pi_j/\pi_r),\qquad j\neq r.$$
An edge $k\to j$ simply indicates that $Z_k^{(r)}$ is a parent of $Z_j^{(r)}$ within this specific coordinate system. It does not strictly imply that ``taxon $k$ affects taxon $j$'' on an absolute scale, which motivates our subsequent reference-node sensitivity analyses to assess skeleton stability. 

While this fixed-reference SEM provides a coherent working target for hypothesis generation, it does not theoretically guarantee causal identifiability, global optimization convergence, or exact false-discovery control; the operational performance of our multinomial layer, truncated penalty, and greedy acyclic projection are therefore evaluated empirically below.

\subsection{Objective Function}
To model directed conditional dependencies among the ALR coordinates, we impose a linear Gaussian structural equation model (SEM) on the latent variables $\bm Z_i$. Let $\mathcal G$ denote a directed acyclic graph (DAG) over nodes $\{1,\dots,p-1\}$ with a weighted adjacency matrix $U=(U_{kj})$. We assume
\begin{equation}
Z_{ij} = \sum_{k \in \mathrm{pa}(j)} U_{kj} Z_{ik} + \varepsilon_{ij}, \qquad \varepsilon_{ij} \sim \mathcal N(0,\sigma_j^2),
\end{equation}
where $\mathrm{pa}(j)$ represents the parent set of node $j$ in $\mathcal G$. The
implemented estimator uses this zero-intercept working SEM and does not center
the latent coordinates before graph fitting. Consequently, fitted coefficients
can reflect both location and conditional-dependence structure; this modeling
restriction is included among the limitations of the real-data analysis. The
SEM is not a mechanistic ecological intervention model.

Furthermore, because purely observational linear Gaussian models are subject to Markov equivalence, true edge orientations cannot be strictly identified without additional assumptions (e.g., non-Gaussian errors or equal variances) which we do not impose. Consequently, the orientations returned by CoLaDAG are algorithm-dependent hypotheses driven by the penalized likelihood and acyclicity constraints; the undirected skeleton and sign stability should be viewed as more robust than any single directional edge.

Under the working independence assumption, we optimize the joint criterion for
the observed counts and sample-specific latent coordinates. This is a joint
optimization criterion rather than a marginal likelihood obtained by
integrating out $\bm Z$. Up to an additive constant, its log-likelihood part is
\begin{align}
\ell(U,\bm\sigma^2,\bm Z)
=&
\sum_{i=1}^n
\left[
\sum_{j=1}^{p-1} X_{ij} Z_{ij}
-
M_i \log\Big(1+\sum_{j=1}^{p-1} \exp(Z_{ij})\Big)
\right]
\nonumber\\
&-
\sum_{j=1}^{p-1}
\left[
\frac{1}{2\sigma_j^2}
\sum_{i=1}^n
\Big(
Z_{ij}
-
\sum_{k\in\mathrm{pa}(j)} U_{kj} Z_{ik}
\Big)^2
+
\frac{n}{2}\log\sigma_j^2
\right].
\label{eq:joint_likelihood}
\end{align}
The multinomial term retains the compositional sampling constraint and the SEM
term supplies a working Gaussian regularizer for the latent coordinates.
Sparsity and acyclicity are imposed in the penalized optimization below.

\subsection{Directed Graph Estimation with Acyclicity Constraint}

Let $d=p-1$ denote the number of modeled latent log-ratio coordinates. The
weighted adjacency matrix $U=(U_{ij})_{d\times d}$ encodes directed SEM coefficients,
with $U_{ij}\neq 0$ indicating an edge from node $i$ to node $j$. We require
this support graph to be acyclic and sparse.

To enforce acyclicity, we use the constrained-likelihood formulation of
\citet{yuan2019}. Introducing auxiliary variables
$\Lambda=(\lambda_{ik})_{d\times d}$, a directed graph is acyclic when its
binary support can be represented through the dual constraints
\begin{equation}
\lambda_{ik} + \mathbb{I}(j\neq k) - \lambda_{jk}
\;\ge\;
\mathbb{I}(U_{ij}\neq 0),
\quad i,j,k=1,\dots,d,\; i\neq j.
\label{eq:dual_acyclic}
\end{equation}
These inequalities rule out directed cycles while preserving the local Markov
interpretation of the DAG \citep{edwards2012}.

The indicator in \eqref{eq:dual_acyclic} is nonconvex and discontinuous. We
therefore follow \citet{shen2012} and use the truncated-$\ell_1$ surrogate
\[
J_\tau(U_{ij}) = \min\left(\frac{|U_{ij}|}{\tau},\,1\right),
\]
where $\tau>0$ controls the approximation. As $\tau\to 0^+$,
$J_\tau(\cdot)$ converges pointwise to the edge indicator. The relaxed
acyclicity constraint becomes
\begin{equation}
\lambda_{ik} + \mathbb{I}(j\neq k) - \lambda_{jk}
\;\ge\;
J_\tau(U_{ij}),
\quad i,j,k=1,\dots,d,\; i\neq j.
\label{eq:relaxed_acyclic}
\end{equation}

The implementation uses a sparsity penalty rather than a fixed edge budget.
Combining the joint likelihood in
\eqref{eq:joint_likelihood} with the relaxed acyclicity constraints gives
\begin{equation}\label{eq:cons_obj}
\begin{aligned}
\text{minimize}\quad
& -\ell(U,\bm\sigma^2,\bm Z)
  +\mu\sum_{i\neq j}J_\tau(U_{ij}) \\
\text{over}\quad
& U,\bm Z,\bm\sigma^2,\Lambda \\
\text{subject to}\quad
& \lambda_{ik} + \mathbb{I}(j\neq k) - \lambda_{jk}
\;\ge\;
J_\tau(U_{ij}),
\quad i,j,k=1,\dots,d,\; i\neq j, \\
& U_{jj}=0,\quad j=1,\dots,d,\qquad \Lambda\geq 0.
\end{aligned}
\end{equation}
Here $\mu>0$ is the sparsity-penalty parameter used by the submitted code. This
formulation aligns with large-scale DAG likelihood approaches
\citep{chun2020} while adapting them to latent compositional coordinates.

\subsection{Final Thresholding and DAG Projection}
\label{subsec:algorithm_threshold}

Following the DC-ADMM optimization, which yields a continuous coefficient matrix $\widehat U^{\mathrm{cont}}$, CoLaDAG applies a hard threshold to determine the final acyclic support. For a final edge threshold $\eta\ge 0$ (which is distinct from the truncated-$\ell_1$ penalty parameter $\tau$), we define the thresholded candidate matrix:
\begin{equation}
\widetilde U_{ij}(\eta) = \widehat U^{\mathrm{cont}}_{ij} \mathbb{I}\{|\widehat U^{\mathrm{cont}}_{ij}|\ge \eta\}, \qquad i\neq j, \quad \widetilde U_{ii}(\eta)=0.
\label{eq:hard_threshold}
\end{equation}
Candidate edges are then sorted in decreasing order of magnitude $|\widetilde U_{ij}(\eta)|$. Starting from an empty graph, CoLaDAG greedily adds a candidate directed edge only if doing so does not create a directed cycle. The final estimator $\widehat U(\eta)$ retains the original signed coefficient for the selected acyclic edges and sets all other entries to zero. Including this explicit truncation step is essential to counteract the densification phenomenon induced by log-ratio transformations.

\section{Simulation Study}\label{sec:simulation}

We evaluated the proposed method in a controlled latent multinomial DAG
benchmark with 100 independent replicates. Each replicate generated a latent
DAG with $d=30$ nodes and $n=500$ samples. Candidate edges were sampled with
probability $2/d$, assigned coefficients $\pm1$, and greedily projected to a
valid DAG. Pre-scaling SEM residual variances were drawn from
$\mathcal U(0.1,0.5)$. The generated coordinates were divided by the softmax
temperature $T=3$ before composition formation; the residual variances on the
final ALR scale were therefore $\sigma_j^2/T^2$. Counts were sampled at fixed
sequencing depth $M=10{,}000$.

We compared CoLaDAG with a project-specific VI-NOTEARS-style implementation,
PC, MMHC, HC-Huge, and Tabu-Huge. CoLaDAG, PC, MMHC, HC-Huge, and Tabu-Huge
received the same latent matrix produced by the preliminary CoLaDAG recovery
wrapper; VI-NOTEARS-style jointly estimated its own latent representation from
the counts. Thus this comparison evaluates graph learners conditional on the
submitted recovery workflow, not six independent end-to-end observation
models. Hyperparameters were fixed using separate tuning simulations rather
than the 100 assessment seeds; the tuning budgets and their asymmetry are given
in Supplementary Section S2. Exact-direction MCC, FDR, and TPR score agreement
with the simulated orientation, whereas SHD is computed after conversion to
CPDAGs. Because the Gaussian observational model does not identify a unique
orientation, CPDAG and skeleton results are the more defensible structural
summaries.

\begin{table}[htbp]
    \centering
    \caption{Main simulation performance over 100 replicated datasets. Entries
    are means with Monte Carlo standard errors in parentheses.}
    \label{tab:simulation_replicates}
    \resizebox{\textwidth}{!}{
    \begin{tabular}{lccccc}
        \toprule
        Method & Edges & MCC & FDR & TPR & SHD \\
        \midrule
        CoLaDAG & $42.32\;(1.171)$ & $0.578\;(0.012)$ & $0.336\;(0.013)$ & $0.546\;(0.009)$ & $36.94\;(1.698)$ \\
        PC & $48.87\;(0.505)$ & $0.374\;(0.009)$ & $0.586\;(0.008)$ & $0.408\;(0.009)$ & $48.53\;(1.080)$ \\
        VI-NOTEARS-style & $32.68\;(0.738)$ & $0.372\;(0.012)$ & $0.501\;(0.014)$ & $0.325\;(0.011)$ & $49.92\;(1.002)$ \\
        Tabu-Huge & $71.74\;(1.098)$ & $0.323\;(0.013)$ & $0.688\;(0.011)$ & $0.440\;(0.013)$ & $65.68\;(1.798)$ \\
        MMHC & $39.22\;(0.421)$ & $0.307\;(0.011)$ & $0.615\;(0.011)$ & $0.306\;(0.010)$ & $48.51\;(1.167)$ \\
        HC-Huge & $71.36\;(1.056)$ & $0.300\;(0.012)$ & $0.706\;(0.010)$ & $0.414\;(0.012)$ & $66.50\;(1.737)$ \\
        \bottomrule
    \end{tabular}
    }
\end{table}

\begin{figure}[htbp]
    \centering
    \includegraphics[width=\textwidth]{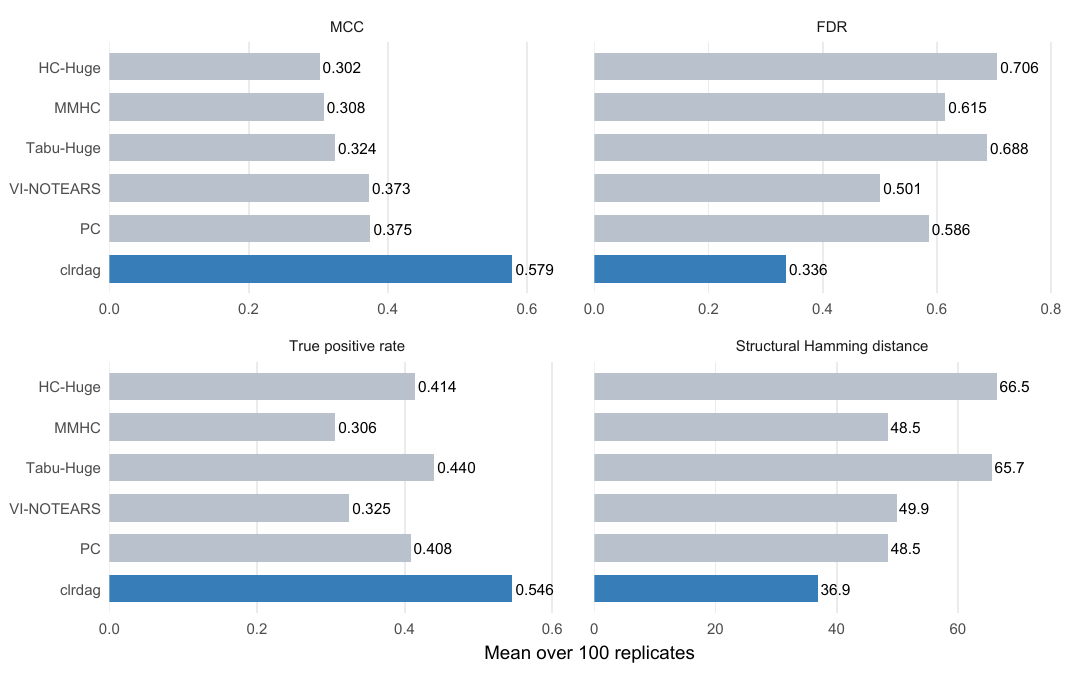}
    \caption{Main simulation performance comparisons.}
    \label{fig:simulation_performance}
\end{figure}

As shown in Table~\ref{tab:simulation_replicates} and
Figure~\ref{fig:simulation_performance}, CoLaDAG had the largest mean
exact-direction MCC and the smallest mean FDR under the aligned latent
multinomial mechanism. PC and the VI-NOTEARS-style implementation followed in
MCC, whereas the graphical-lasso-screened HC and Tabu searches produced denser
graphs and larger CPDAG SHD. These results describe finite-sample behavior under
the stated generating mechanism; they do not establish orientation
identifiability or uniform superiority.

\subsection{Sensitivity Analysis for Log-ratio Sparsity}
\label{subsec:logratio_sparsity_sim}

To illustrate the non-invariance concerns in
Section~\ref{subsec:logratio_graph}, we conducted a population-level diagnostic
independent of any estimator. Sparse linear Gaussian DAGs were generated under
chain, hub, and Erdos--Renyi topologies with absolute-scale dimension
$q\in\{20,50,100\}$. We transformed the exact population covariance into ALR
coordinates and into an orthonormal Helmert (ILR) basis, then evaluated
precision and order-specific DAG densities at numerical tolerance $10^{-8}$.

\begin{table}[htbp]
    \centering
    \caption{Log-ratio sparsity diagnostic at absolute-scale dimension $q=50$.}
    \label{tab:logratio_sparsity_diagnostic}
    \small
    \setlength{\tabcolsep}{2pt}
    \begin{tabular}{@{}llcccc@{}}
        \toprule
        Topology & Transform & \shortstack{Original DAG\\density} &
        \shortstack{Precision\\density} & \shortstack{Ordered DAG\\density} &
        \shortstack{Random-order\\DAG density} \\
        \midrule
        Chain & ALR & $0.040$ & $1.000$ & $1.000$ & $1.000$ \\
        Hub & ALR & $0.060$ & $1.000$ & $1.000$ & $1.000$ \\
        Random & ALR & $0.039$ & $1.000$ & $1.000$ & $1.000$ \\
        Chain & ILR (Helmert) & $0.040$ & $1.000$ & $1.000$ & $1.000$ \\
        Hub & ILR (Helmert) & $0.060$ & $1.000$ & $1.000$ & $1.000$ \\
        Random & ILR (Helmert) & $0.039$ & $1.000$ & $1.000$ & $1.000$ \\
        \bottomrule
    \end{tabular}
\end{table}

At the stated tolerance, the transformed precision matrices and sequential DAG
factorizations were complete in the reported settings
(Table~\ref{tab:logratio_sparsity_diagnostic}). This diagnostic illustrates the
potential for severe sparsity loss under the examined references, bases, and
coefficient distributions. It motivates sensitivity analysis and explicit
thresholding, but does not prove that every log-ratio graph is dense or validate
a particular threshold.

\subsection{Extended Robustness and Ablation Checks}
\label{subsec:extended_simulations}

Supplementary Section S3 reports component ablation, dimension and graph-density
experiments, continuous-SEM misspecification, random count dropout, and final
threshold sensitivity. The aligned experiments favored CoLaDAG on mean MCC,
whereas HC-Huge was substantially better for directly observed continuous SEM
data and all methods were near chance after the examined dropout contamination.
The supplement therefore maps model-specific operating characteristics rather
than confirming a universal method ranking.

\section{Real Data Analysis}
\label{sec:realdata}

\subsection{Genus-level Graph and Biological Interpretability}
\label{subsec:real_genus_network}
To derive biological insights from the real data, we applied CoLaDAG to the
genus abundance table, using the most abundant retained genus,
Lachnospiraceae\_NK4A136\_group, as the ALR reference denominator. CoLaDAG
identified 284 directed ALR-coordinate edges, whereas the baseline methods
selected between 34 and 89 edges under their specified thresholds
(Table~\ref{tab:real_data_edges}). Given the density of the full support graph,
biological interpretation prioritizes the strongest coefficients, supplemented
by time-adjusted ALR slope comparisons, mouse-block stability, and
reference-node sensitivity analyses.

\begin{table}[htbp]
    \centering
    \caption{Primary genus-level edge counts.}
    \label{tab:real_data_edges}
    \begin{tabular}{lcc}
        \toprule
        Method & Discovered Edges & Network Sparsity \\
        \midrule
        CoLaDAG (Ours) & $284$ & $8.59\%$ \\
        HC-Huge & $89$ & $2.69\%$ \\
        PC & $53$ & $1.60\%$ \\
        MMHC & $47$ & $1.42\%$ \\
        Tabu-Huge & $89$ & $2.69\%$ \\
        VI-NOTEARS-style & $34$ & $1.03\%$ \\
        \bottomrule
    \end{tabular}
\end{table}

The top 10 strongest CoLaDAG coefficients, representing fixed-reference ALR-coordinate dependencies within the fitted SEM, are prioritized in Table \ref{tab:top_CoLaDAG_edges} for biological interpretation.

\begin{table}[htbp]
    \centering
    \caption{Top primary CoLaDAG edges.}
    \label{tab:top_CoLaDAG_edges}
    \small
    \setlength{\tabcolsep}{3pt}
    \begin{tabular}{@{}rP{0.23\textwidth}P{0.30\textwidth}rrl@{}}
        \toprule
        Rank & From & To & Weight & $|\mathrm{Weight}|$ & Sign \\
        \midrule
        1 & Intestinimonas & Lachnospiraceae\_\allowbreak AC2044\_\allowbreak group & 3.5196 & 3.5196 & Positive \\
        2 & Alistipes & Ruminiclostridium\_\allowbreak 5 & 2.6899 & 2.6899 & Positive \\
        3 & Oscillibacter & Rikenellaceae\_\allowbreak RC9\_\allowbreak gut\_\allowbreak group & 2.6810 & 2.6810 & Positive \\
        4 & Oscillibacter & Lachnospiraceae\_\allowbreak AC2044\_\allowbreak group & -2.3723 & 2.3723 & Negative \\
        5 & Intestinimonas & Azospirillum\_\allowbreak sp.\_\allowbreak 47\_\allowbreak 25 & 2.0719 & 2.0719 & Positive \\
        6 & Alistipes & Dubosiella & -1.9843 & 1.9843 & Negative \\
        7 & Intestinimonas & Dubosiella & -1.8766 & 1.8766 & Negative \\
        8 & Ruminiclostridium & Parabacteroides & -1.8211 & 1.8211 & Negative \\
        9 & Lachnoclostridium & A2 & 1.7584 & 1.7584 & Positive \\
        10 & Roseburia & Faecalibaculum & 1.6581 & 1.6581 & Positive \\
        \bottomrule
    \end{tabular}
\end{table}

The first-ranked relation was from Intestinimonas to the Lachnospiraceae
AC2044 group, a database-defined genus-level feature. It attained mouse-block
selection frequency 0.70 and conditional sign agreement 0.857, providing
stronger exploratory support than rank alone. The exact labels Oscillibacter
and Rikenellaceae RC9 gut group were verified against the supplied genus table.
Their fitted relation may reflect coordinated change, but the present
compositional analysis cannot distinguish shared responses to diet, exposure,
or host state from microbial exchange or mathematical coupling induced by the
common ALR denominator. It should therefore not be interpreted as evidence of
direct ecological interaction or cross-feeding. Follow-up studies should
combine taxon-specific quantification with SCFA and other metabolite
measurements.

\subsection{Mouse-block Stability Analyses}
\label{subsec:bootstrap_stability}

Because the real AgNP data do not have a known ground-truth network, baseline
parameters cannot be selected by MCC or FDR. We therefore performed an
additional mouse-block bootstrap stability experiment for the score-based
HC-Huge and Tabu-Huge baselines. The experiment used the 30 ALR coordinates with
the largest variance under the same primary reference taxon, evaluated StARS
thresholds $\{0.10,0.15,0.20\}$ and BIC penalty multipliers $\{1,2,3\}$, and
repeated each setting over 25 bootstrap resamples. Each bootstrap replicate
sampled the 12 mouse blocks with replacement and retained all time points for a
selected mouse. An edge was treated as stable if its empirical selection frequency was
at least 0.60.
We summarized each setting by the mean number of skeleton edges, mean number of
directed graph edges, number of stable directed edges, and an instability proxy
$\overline{2\pi_e(1-\pi_e)}$, where $\pi_e$ is the bootstrap selection
probability of edge $e$.

The most stable configuration was Tabu-Huge with StARS threshold 0.10 and BIC
multiplier 3, yielding a mean of 14.16 directed edges, 4 stable edges, and the
lowest instability score (0.0222). The corresponding HC-Huge setting selected
15.80 directed edges on average and 3 stable edges. Compared with ordinary
sample-level resampling, mouse-block bootstrap gives a more conservative view of
edge stability and reinforces the need to treat all real-data graphs as
exploratory.

\begin{table}[htbp]
    \centering
    \caption{Mouse-block bootstrap baseline tuning.}
    \label{tab:bootstrap_stability}
    \small
    \setlength{\tabcolsep}{2.5pt}
    \begin{tabular}{@{}lrrrrrrr@{}}
        \toprule
        Method & StARS & $k$ factor & \shortstack{Bootstrap\\reps.} &
        \shortstack{Mean\\skeleton\\edges} & \shortstack{Mean\\graph\\edges} &
        \shortstack{Stable\\edges} & Instability \\
        \midrule
        Tabu-Huge & 0.10 & 3 & 25 & 28.56 & 14.16 & 4 & 0.0222 \\
        HC-Huge & 0.10 & 3 & 25 & 33.60 & 15.80 & 3 & 0.0242 \\
        Tabu-Huge & 0.10 & 2 & 25 & 33.96 & 16.36 & 4 & 0.0253 \\
        HC-Huge & 0.10 & 2 & 25 & 34.32 & 16.80 & 3 & 0.0264 \\
        HC-Huge & 0.10 & 1 & 25 & 31.24 & 18.76 & 6 & 0.0272 \\
        Tabu-Huge & 0.15 & 3 & 25 & 76.64 & 18.32 & 5 & 0.0280 \\
        \bottomrule
    \end{tabular}
\end{table}

We also performed a reduced-iteration CoLaDAG mouse-block diagnostic. Twenty
resamples drew the 12 mouse blocks with replacement and retained all four
observations from each selected block. The diagnostic used smaller DC and ADMM
iteration caps than the primary fit for computational feasibility and therefore
does not estimate the stability of an identical refitting procedure. All 20
scripts returned a graph, although the current wrapper does not expose
inner-solver fallback counts. The resampled graphs contained 603.15 edges on
average, compared with 284 in the primary fit, revealing substantial tuning and
resampling sensitivity. Among the 284 primary edges, the median selection
frequency was 0.45; 60 edges had frequency at least 0.60, and 37 also had
conditional sign agreement at least 0.80. The leading primary
Intestinimonas $\to$ Lachnospiraceae\_AC2044\_group edge had selection
frequency 0.70 and conditional sign agreement 0.857. These frequencies are
descriptive stability diagnostics, not selection-adjusted error probabilities.

\begin{table}[htbp]
    \centering
    \caption{Reduced-iteration CoLaDAG mouse-block stability diagnostic.}
    \label{tab:CoLaDAG_block_stability}
    \begin{tabular}{@{}lr@{}}
        \toprule
        Quantity & Value \\
        \midrule
        Successful bootstrap fits & 20 / 20 \\
        Primary edges & 284 \\
        Mean bootstrap edges & 603.15 \\
        Median primary-edge selection frequency & 0.45 \\
        Primary edges with selection frequency $\geq 0.60$ & 60 \\
        Above threshold with sign agreement $\geq 0.80$ & 37 \\
        \bottomrule
    \end{tabular}
\end{table}

\subsection{Dose-stratified Analyses}
\label{subsec:dose_network}

The AgNP design naturally motivates comparison across the control, 0.1Ag, 2Ag,
and 40Ag groups. Each exposure group contains only three mice and 12
longitudinal observations, so relearning a separate 58-node DAG in each group
would be unstable. We therefore used a restricted shared-support display. The
global CoLaDAG fit supplied candidate orientations, after which only its 80
largest absolute coefficients were considered. Within each exposure group, we
fitted a separate regression of the child ALR coordinate on one candidate
parent and collection-time indicators. These quantities are time-adjusted
pairwise ALR slopes, not conditional SEM coefficients, because the remaining
parents of the child were not included. Slopes with
$|\widehat w|\geq 0.75$ were retained and at most 20 were displayed per group.
The threshold and top-20 cap are visualization rules, not dose-specific DAG
tuning parameters. Displayed arrows retain the global orientation solely to
identify the source and target coordinates.

\begin{figure}[htbp]
    \centering
    \includegraphics[width=\textwidth]{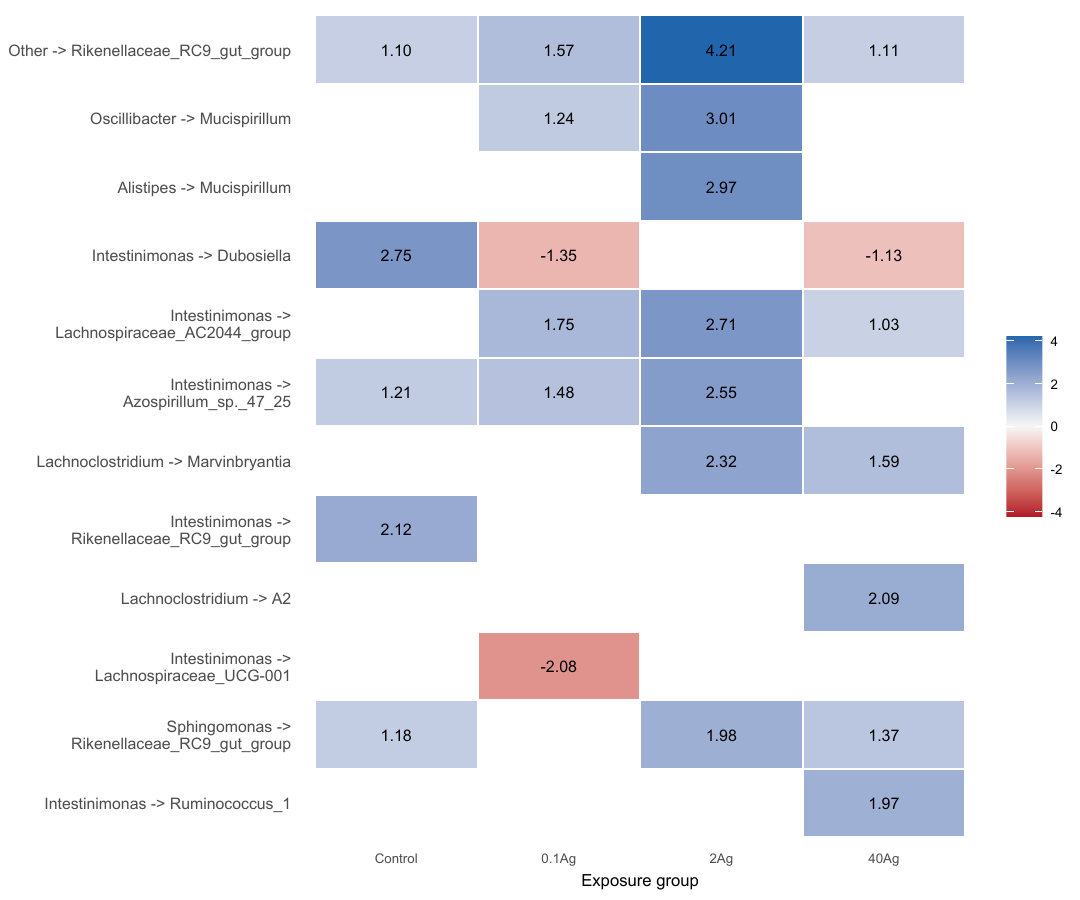}
    \caption{Time-adjusted pairwise ALR slopes for the retained dose-overlay
    relations. Slopes are descriptive and preserve the global orientation only
    as a labeling convention.}
    \label{fig:dose_group_heatmap}
\end{figure}

\begin{figure}[htbp]
    \centering
    \includegraphics[width=\textwidth]{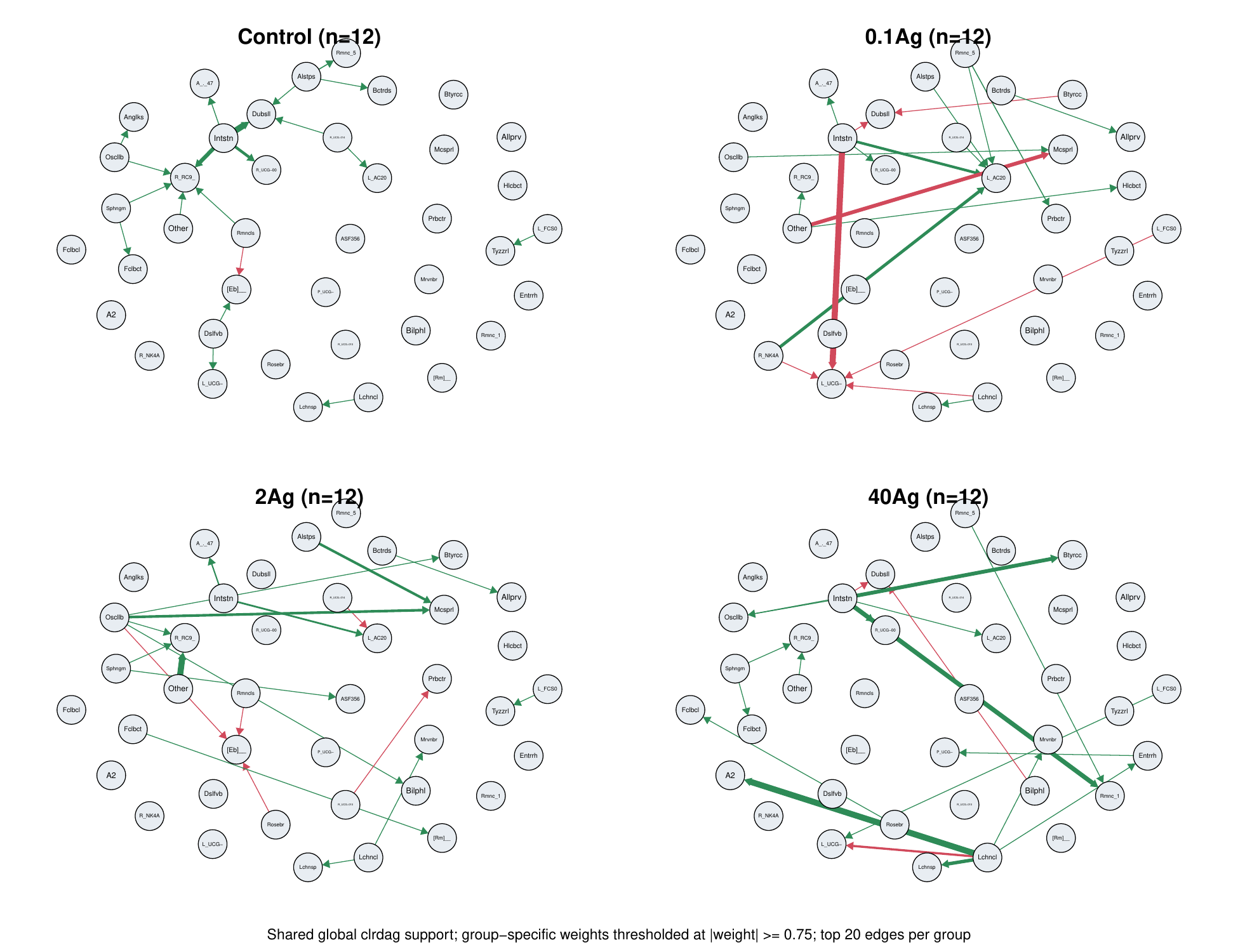}
    \caption{Thresholded exposure-group displays of time-adjusted pairwise ALR
    slopes on restricted global CoLaDAG candidate orientations.}
    \label{fig:dose_group_graphs}
\end{figure}

These displays do not represent four independently estimated DAGs. The heatmap
is the primary comparison; the network panels retain the global source--target
labels for context. Slopes that remain large in several groups suggest a
recurrent pairwise log-ratio pattern, whereas sign or magnitude differences are
descriptive candidates for a future mouse-level interaction analysis. The full
ranked lists are reported in Supplementary Section S4 and in the accompanying
CSV file.

The primary denominator was the Lachnospiraceae NK4A136 group. Under this
reference, the Intestinimonas-to-Dubosiella slope was 2.752 in Control,
$-1.350$ in 0.1Ag, 0.076
in 2Ag, and $-1.132$ in 40Ag. The Alistipes--Ruminiclostridium group 5 slope was
positive in all four groups (1.261, 0.516, 1.013, and 0.886, respectively). The
first pattern illustrates heterogeneity, whereas the second illustrates a more
recurrent relation; neither establishes a dose response. Each group contains
only three mice, and the naive intervals ignore support selection,
within-mouse dependence, and multiplicity. We therefore report these slopes as
a biological follow-up list rather than as significant exposure-specific
effects. The pooled ``Other'' category is excluded from biological
interpretation because it is not a taxon.

\section{Discussion and Conclusion}
\label{sec:discussion}

Under simulations aligned with its multinomial--latent construction, CoLaDAG
achieved the largest mean exact-direction MCC and the smallest mean FDR among
the evaluated implementations. The advantage did not generalize to every
setting: HC-Huge performed better for the unaligned continuous SEM, and random
count dropout reduced all methods to near-chance recovery. The simulation
evidence is therefore model-specific rather than a universal ranking.

In the AgNP study, the genus-level graph prioritized coordinate-specific
hypotheses involving Intestinimonas, Alistipes, Oscillibacter,
Ruminiclostridium group 5, and Lachnospiraceae AC2044 group. Only 60 of 284
primary edges reached mouse-block selection frequency 0.60, 37 also met the
sign-agreement criterion, and alternative ALR references produced materially
different graphs. These findings support a ranked follow-up list, not stable or
reference-invariant microbial interactions.

Time-adjusted pairwise ALR slopes showed both heterogeneous and recurrent
patterns across exposure groups, but three mice per group cannot establish dose
response or exposure effects. Together with the zero-intercept working SEM,
nonconvex optimization, repeated-measure treatment in the primary likelihood,
and unresolved genus-group labels, these limitations make CoLaDAG a
fixed-reference screening procedure for directed log-ratio hypotheses rather
than a microbial causal-map estimator. Independent abundance, metabolite, and
perturbation studies are required for biological validation.

\section*{Declaration of Generative AI and AI-assisted Technologies}

During preparation of this work, the authors used OpenAI Codex to support
language editing, code auditing, and manuscript formatting. The authors reviewed
and edited all resulting material and take full responsibility for the content
of the publication.

\section*{Funding}
Shuyan Chen was supported by the Youth Innovation Fund of the University of
Science and Technology of China (No. WK2040250119).

\bibliographystyle{elsarticle-harv}
\bibliography{Bibliography}

\clearpage
\setcounter{section}{0}
\setcounter{table}{0}
\setcounter{figure}{0}
\setcounter{equation}{0}
\renewcommand{\thesection}{S\arabic{section}}
\renewcommand{\thetable}{S\arabic{table}}
\renewcommand{\thefigure}{S\arabic{figure}}
\renewcommand{\theequation}{S\arabic{equation}}

\begin{center}
{\large Supplementary Material for}\par
\vspace{0.4em}
{\Large CoLaDAG: Compositional Latent Log-ratio DAG Analysis of the Gut
Microbiome under Silver Nanoparticle Exposure}\par
\vspace{0.7em}
Shuyan Chen, Ziliang Shen, and Xinlei Wang
\end{center}

\section{Interpretation of Taxonomic Labels}
\label{sec:supp_taxa}

The real-data tables and network figures preserve the machine-readable labels
from the supplied genus-level count table. Underscores are separators and do
not indicate additional taxonomic resolution. The interpretations below decode
the labels used in the displays; they are not species-level reannotations.

\begin{small}
\begin{longtable}{@{}P{0.28\textwidth}P{0.23\textwidth}P{0.41\textwidth}@{}}
\caption{Nonstandard taxonomic labels.}
\label{tab:supp_taxon_labels}\\
\toprule
Original label & Readable form & Interpretation and reporting level \\
\midrule
\endfirsthead
\toprule
Original label & Readable form & Interpretation and reporting level \\
\midrule
\endhead
\bottomrule
\endfoot
\path{Lachnospiraceae_NK4A136_group} & Lachnospiraceae NK4A136 group & Database-defined group within Lachnospiraceae; genus-level group label and the primary ALR denominator. \\
\path{Rikenellaceae_RC9_gut_group} & Rikenellaceae RC9 gut group & Database-defined RC9 gut group within Rikenellaceae; genus-level group label, not a confirmed species. \\
\path{Lachnospiraceae_AC2044_group} & Lachnospiraceae AC2044 group & Database-defined AC2044 group within Lachnospiraceae; genus-level group label. \\
\path{Lachnospiraceae_UCG-001} & Lachnospiraceae UCG-001 & Database-defined uncultured genus-level cluster within Lachnospiraceae. \\
\path{Ruminococcaceae_UCG-004} & Ruminococcaceae UCG-004 & Database-defined uncultured genus-level cluster within Ruminococcaceae. \\
\path{Ruminococcaceae_UCG-010} & Ruminococcaceae UCG-010 & Database-defined uncultured genus-level cluster within Ruminococcaceae. \\
\path{Ruminococcaceae_UCG-014} & Ruminococcaceae UCG-014 & Database-defined uncultured genus-level cluster within Ruminococcaceae. \\
\path{Ruminococcaceae_NK4A214_group} & Ruminococcaceae NK4A214 group & Database-defined NK4A214 group within Ruminococcaceae; genus-level group label. \\
\path{[Eubacterium]_xylanophilum_group} & Eubacterium xylanophilum group & Bracketed legacy genus assignment associated with the xylanophilum group; not a species-level identification. \\
\path{[Ruminococcus]_gnavus_group} & Ruminococcus gnavus group & Bracketed legacy genus assignment associated with the gnavus group; not a species-level identification. \\
\path{Ruminiclostridium_5} & Ruminiclostridium group 5 & Numbered database subgroup within Ruminiclostridium; not a formal species name. \\
\path{Ruminiclostridium_6} & Ruminiclostridium group 6 & Numbered database subgroup within Ruminiclostridium; not a formal species name. \\
\path{Ruminococcus_1} & Ruminococcus group 1 & Numbered database subgroup within Ruminococcus; genus-level feature label. \\
\path{Azospirillum_sp._47_25} & Azospirillum sp. 47-25 & Supplied feature identifier; the present data do not independently validate a species or strain. \\
\path{A2} & A2 & Unresolved database feature label retained from the supplied taxonomy table. \\
\path{ASF356} & ASF356 & Altered Schaedler Flora member identifier; a strain/group label rather than a genus name. \\
\path{Other} & Other & Pooled residual category; not a single biological taxon and not used for mechanistic interpretation. \\
\end{longtable}
\end{small}

\section{Computational Implementation and Comparator Configuration}
\label{sec:supp_implementation}

\subsection{Implemented target, initialization, and stopping rules}

The submitted workflow optimizes the joint count--latent-coordinate criterion
given in the main manuscript. It is a joint optimization over the
sample-specific $\bm Z_i$ rather than a marginal likelihood. The implemented
working SEM has no intercept and the latent coordinates are not centered before
graph fitting. This restriction is disclosed because it can cause coefficients
to absorb coordinate location as well as conditional-dependence structure.

For a count matrix with $n$ rows and $p$ parts, $d=p-1$ coordinates are modeled.
The default initialization is
\[
Z_{ij}^{(0)}=\log\frac{X_{ij}+0.5}{X_{ir}+0.5},\qquad
U^{(0)}=0,\qquad \Lambda^{(0)}=0,
\]
where $r$ is the final, prespecified reference column. The initial residual
variances are the empirical coordinate variances, truncated below at 0.05. Raw
counts and sample totals enter the multinomial likelihood after initialization;
the 0.5 pseudocount is also used to construct continuous ALR inputs for baseline
methods.

Each sample-specific latent vector is updated by BFGS with an analytic gradient.
The outer wrapper stops when the relative change in its recorded objective is
below $10^{-3}$ or the configured cap is reached. The graph solver receives
absolute and relative ADMM tolerances of $10^{-3}$. The analysis scripts also
apply fixed DC and ADMM caps, so reaching a cap is not equivalent to a
convergence certificate. The optimization is nonconvex and no global optimum is
claimed.

\subsection{Parameter configurations and tuning provenance}

\begin{table}[htbp]
\centering
\caption{Estimator inputs, tuning, and directed-output conventions.}
\label{tab:supp_comparators}
\small
\setlength{\tabcolsep}{3pt}
\begin{tabular}{@{}P{0.15\textwidth}P{0.18\textwidth}P{0.40\textwidth}P{0.18\textwidth}@{}}
\toprule
Method & Input & Configuration & Output convention \\
\midrule
CoLaDAG, simulation & Counts and recovered latent ALR coordinates & Recovery: $\tau=0.10$, $\mu=4$, 3 DC and 20 ADMM iterations. Refinement: $\tau=0.25$, $\mu=1$, $\rho=1.5$, 10 DC and 50 ADMM iterations; final threshold $\eta=0.30$. & $U_{ij}\ne0$ means $i\to j$ \\
CoLaDAG, real data & Counts and recovered latent ALR coordinates & Recovery: $\tau=0.10$, $\mu=2$, 5 DC and 20 ADMM iterations. Refinement: $\tau=0.08$, $\mu=1$, $\rho=1.5$, 10 DC and 50 ADMM iterations; final threshold $\eta=0.08$. & $U_{ij}\ne0$ means $i\to j$ \\
HC-Huge / Tabu-Huge & Shared recovered ALR matrix & Graphical-lasso skeleton selected by StARS 0.20; Gaussian BIC score with $k=\log n$; absent skeleton arcs blacklisted. & \texttt{bnlearn} \texttt{from}$\to$\texttt{to} converted without transposition \\
PC & Shared recovered ALR matrix & Gaussian correlation tests with $\alpha=0.05$; default stable-PC options. & Returned arcs converted to the common adjacency convention \\
MMHC & Shared recovered ALR matrix & Default \texttt{bnlearn::mmhc} configuration. & Returned arcs converted without transposition \\
VI-NOTEARS-style & Multinomial counts & Project-specific variational implementation; $\lambda=2$, 10 outer iterations, 500 inner iterations, final threshold 0.50, and greedy DAG projection. & Row source, column target \\
\bottomrule
\end{tabular}
\end{table}

The HC-Huge and Tabu-Huge StARS/BIC choices were examined in a truth-based grid
at $n=100,d=30$ using two tuning seeds. The VI-NOTEARS-style value of $\lambda$
and its threshold were selected on one separate default simulation. PC and MMHC
used the stated defaults. These tuning budgets are not symmetric; the manuscript
therefore reports a comparison of the submitted implementations rather than a
method-independent ranking. None of the 100 primary assessment seeds was used
to select these settings. Real-data parameters were not selected using an
unknown graph truth. The threshold $\eta=0.08$ is an operational screening
choice and is not presented as an error-controlling cutoff.

Directed MCC, FDR, and TPR score the exact simulated orientation. SHD is
evaluated after converting the estimated and simulated DAGs to CPDAGs. The unit
test constructs $V_1\to V_2\to V_3$ and checks metric calculation, graph
conversion, edge-list writing, and plotting. Correcting an earlier unintended
transpose in the \texttt{bnlearn} conversion required regeneration of the
affected outputs.

\subsection{Optimization blocks and complexity}

For $d=p-1$, the negative multinomial component is
\begin{equation}
-\ell_{\mathrm{mn}}(\bm Z)=
-\sum_{i=1}^n\left[
\sum_{j=1}^d X_{ij}Z_{ij}
-M_i\log\left(1+\sum_{j=1}^d e^{Z_{ij}}\right)
\right],
\end{equation}
and the zero-intercept SEM component is
\begin{equation}
-\ell_{\mathrm{sem}}(U\mid\bm Z,\bm\sigma^2)=
\sum_{j=1}^d\left[
\frac{1}{2\sigma_j^2}\sum_{i=1}^n
\left(Z_{ij}-\sum_{k\ne j}U_{kj}Z_{ik}\right)^2
+\frac{n}{2}\log\sigma_j^2
\right].
\end{equation}
Given $(U,\bm\sigma^2)$, BFGS minimizes the smooth joint objective over each
$\bm Z_i$. Given $(\bm Z,U)$, the residual-variance update is
\[
\sigma_j^2=\max\left\{0.05,\frac{1}{n}\sum_{i=1}^n
\left(Z_{ij}-\sum_{k\ne j}U_{kj}Z_{ik}\right)^2\right\}.
\]
The truncated surrogate
\[
J_\tau(z)=\min(|z|/\tau,1)
=|z|/\tau-\max(|z|/\tau-1,0)
\]
is handled by difference-of-convex iterations. ADMM alternates nodewise ridge
updates for $U$, weighted soft thresholding for the split coefficient matrix,
least-squares updates for nonnegative ordering variables, projection of
acyclicity slacks onto the nonnegative orthant, and scaled-dual updates. The
continuous coefficient matrix is finally thresholded at $\eta$ and greedily
projected to a DAG in decreasing absolute-coefficient order.

Let $L_B$ be the number of BFGS evaluations. A dense outer latent sweep costs
$O(nL_Bd^2+nd^2+d^3)$ time and $O(nd+d^2)$ memory. The direct ADMM formulation
stores slack and dual arrays indexed by $(i,j,k)$ and therefore requires
$O(d^3)$ working memory. Its $d$ ordering-variable least-squares problems have
$O(d^2)$ rows and $d$ columns; without cached factorizations this gives a
conservative $O(d^5)$ time bound per ADMM sweep. Nodewise ridge solves add at
most $O(nd^2+d^4)$ time. These are implementation bounds, not claims of optimal
algorithmic complexity.

\subsection{Initialization sensitivity and numerical reporting}

One default and nine jittered starts were evaluated in an aligned $n=300,d=20$
simulation. Jittered latent and graph initial values used standard deviation
0.10.

\begin{table}[htbp]
\centering
\caption{Initialization sensitivity of CoLaDAG.}
\label{tab:supp_initialization}
\small
\begin{tabular}{@{}rrrrr@{}}
\toprule
Start & Edges & MCC & Jaccard to default & Final objective \\
\midrule
1 & 22 & 0.577 & 1.000 & 7,881,881.391 \\
2 & 23 & 0.604 & 0.800 & 7,881,880.782 \\
3 & 24 & 0.589 & 0.769 & 7,882,138.519 \\
4 & 23 & 0.604 & 0.800 & 7,881,904.771 \\
5 & 23 & 0.604 & 0.800 & 7,881,952.590 \\
6 & 20 & 0.745 & 0.680 & 7,882,071.531 \\
7 & 24 & 0.589 & 0.769 & 7,882,214.882 \\
8 & 25 & 0.575 & 0.741 & 7,882,129.516 \\
9 & 24 & 0.589 & 0.769 & 7,882,214.223 \\
10 & 23 & 0.604 & 0.800 & 7,881,895.727 \\
\bottomrule
\end{tabular}
\end{table}

Support Jaccard similarity ranged from 0.680 to 1.000 and MCC from 0.575 to
0.745, indicating moderate local-solution sensitivity. Every outer wrapper in
this diagnostic stopped after two iterations. The current scripts catch some
inner-solver errors and can return the previous graph; therefore a returned graph
must not automatically be interpreted as a numerically converged fit.

\subsection{Computing environment and archive}

The archived session records R 4.4.2 on 64-bit Windows 11 (build 26200). The
workstation is a Lenovo 82JU with an AMD Ryzen 7 5800H processor (8 physical, 16
logical cores; nominal maximum clock 3.2 GHz) and 15.9 GB RAM. Analyses were run
without a documented distributed-computing backend. The current session file
does not capture the complete contributed-package version set; this is a
reproducibility limitation that should be remedied with a lockfile in any public
release.

\begin{sloppypar}
The accompanying reproducibility ZIP contains the manuscript-facing R scripts,
processed toy/analysis data, simulation summaries, real-data edge tables,
figures, fixed seeds, and the entry point
\path{R/00_run_submission_pipeline.R}. The latest archived driver status
\texttt{verified\_existing} verifies packaged outputs but does not claim that all
analyses were rerun in that invocation. A clean rerun should preserve solver
logs and fail rather than silently accept an inner-solver fallback.
\end{sloppypar}

\subsection{Implementation and Optimization Strategy}
\label{subsec:tuning}

The submitted implementation initializes $Z$ with pseudocount ALR coordinates,
$U$ and $\Lambda$ with zero matrices, and residual variances with the empirical
coordinate variances truncated below at 0.05. Conditional on $(U,\bm\sigma^2)$,
each sample-specific latent vector is updated by the
Broyden--Fletcher--Goldfarb--Shanno (BFGS) algorithm using an analytic gradient.
Residual variances are then updated from the nodewise squared residuals, followed
by DC and ADMM graph updates. The outer wrapper stops when the relative change in
its recorded objective is below $10^{-3}$ or its iteration limit is reached; the
graph solver receives absolute and relative ADMM tolerances of $10^{-3}$. The
overall problem is nonconvex. In the 10-start diagnostic, support Jaccard
similarity to the default start ranged from 0.680 to 1.000 and exact-direction
MCC ranged from 0.575 to 0.745 after excluding diagonal entries from the
confusion matrix, indicating moderate rather than negligible
initialization sensitivity.

For dense matrices, one outer latent-coordinate sweep requires
$O(nL_Bd^2+nd^2+d^3)$ operations and $O(nd+d^2)$ storage, where $L_B$ is the
number of BFGS objective/gradient evaluations. The implemented ADMM graph block
uses $O(d^3)$ acyclicity slack and dual arrays. Its direct ordering-variable
least-squares updates give a conservative worst-case bound of $O(d^5)$ time per
ADMM sweep and $O(d^3)$ working memory; the nodewise ridge updates contribute at
most $O(nd^2+d^4)$ additional time. These bounds explain why the $d=58$,
12-mouse analysis is treated as a heavily regularized screening calculation.

All parameter values, iteration caps, comparator settings, initialization
diagnostics, and reproducibility instructions are reported in Supplementary
Section S2. Because the effective biological sample size is 12, neither the
tuning parameters nor $\eta$ turn the case study into confirmatory
high-dimensional graph inference.

\section{Extended Simulations and Robustness Checks}
\label{sec:supp_extended}

\subsection{Sample size and sequencing depth}

The aligned sensitivity experiments used $d=20$ and 10 independent replicates
per setting. The table reports CoLaDAG and the strongest comparator by mean MCC;
complete method-by-setting rows are included in the CSV archive.

\begin{table}[htbp]
\centering
\caption{Sample-size and sequencing-depth sensitivity.}
\label{tab:supp_sample_depth}
\small
\resizebox{\textwidth}{!}{%
\begin{tabular}{@{}lcccc@{}}
\toprule
Setting & CoLaDAG MCC & Best comparator & Comparator MCC & CoLaDAG FDR \\
\midrule
$n=100$ & $0.434\;(0.054)$ & PC & $0.351\;(0.040)$ & 0.520 \\
$n=300$ & $0.417\;(0.040)$ & PC & $0.428\;(0.039)$ & 0.509 \\
$n=500$ & $0.509\;(0.066)$ & PC & $0.406\;(0.034)$ & 0.410 \\
$M=1{,}000$ & $0.328\;(0.053)$ & PC & $0.249\;(0.018)$ & 0.564 \\
$M=5{,}000$ & $0.470\;(0.059)$ & PC & $0.375\;(0.034)$ & 0.442 \\
$M=10{,}000$ & $0.548\;(0.042)$ & Tabu-Huge & $0.390\;(0.042)$ & 0.369 \\
\bottomrule
\end{tabular}
}
\end{table}

CoLaDAG was not uniformly best: PC had a slightly larger mean MCC at $n=300$.
The results describe operating characteristics near the aligned benchmark rather
than a fixed method ordering.

\subsection{Unaligned continuous SEM and dropout}

The continuous stress test generated directly observed SEM variables with
Gaussian, heavy-tailed $t_3$, or centered exponential errors. There was no
multinomial observation layer. Each setting used $n=300,d=20$ and 20 replicates.

\begin{table}[htbp]
\centering
\caption{Unaligned continuous-SEM stress test.}
\label{tab:supp_unaligned}
\small
\begin{tabular}{@{}llrrr@{}}
\toprule
Error & Method & Edges & MCC & FDR \\
\midrule
Gaussian & HC-Huge & 14.75 & $0.808\;(0.024)$ & 0.093 \\
Gaussian & CoLaDAG & 6.25 & $0.241\;(0.046)$ & 0.394 \\
Heavy-tailed $t_3$ & HC-Huge & 14.35 & $0.761\;(0.025)$ & 0.143 \\
Heavy-tailed $t_3$ & CoLaDAG & 7.00 & $0.243\;(0.033)$ & 0.452 \\
Centered exponential & HC-Huge & 15.30 & $0.761\;(0.019)$ & 0.150 \\
Centered exponential & CoLaDAG & 7.05 & $0.317\;(0.059)$ & 0.380 \\
\bottomrule
\end{tabular}
\end{table}

The score-based baseline was substantially better in this unaligned setting.
This negative result prevents generalizing the aligned-model advantage to
directly observed continuous SEM data.

After multinomial sampling, the dropout experiment independently set observed
nonzero counts to zero with probability 0, 0.10, or 0.30. Each setting used
$n=300,d=20,M=10{,}000$ and 10 replicates.

\begin{table}[htbp]
\centering
\caption{Random count-dropout stress test.}
\label{tab:supp_dropout}
\small
\begin{tabular}{@{}llrrr@{}}
\toprule
Dropout & Method & Edges & MCC & FDR \\
\midrule
0 & CoLaDAG & 28.7 & $0.477\;(0.035)$ & 0.424 \\
0 & PC & 28.4 & $0.425\;(0.023)$ & 0.480 \\
0.10 & CoLaDAG & 6.1 & $-0.008\;(0.014)$ & 0.925 \\
0.10 & PC & 45.9 & $-0.012\;(0.009)$ & 0.922 \\
0.30 & CoLaDAG & 10.7 & $0.014\;(0.016)$ & 0.893 \\
0.30 & PC & 47.9 & $-0.009\;(0.015)$ & 0.925 \\
\bottomrule
\end{tabular}
\end{table}

All evaluated methods were near chance after 10\% or 30\% dropout. The present
observation model has no structural-zero or zero-inflation component.

\subsection{Component ablation}

The ablation used 30 replicates with $n=300,d=20,M=10{,}000$. The oracle fit
received simulated latent coordinates and is an unattainable reference. The
observed-ALR fit used pseudocount log ratios without latent recovery.

\begin{table}[htbp]
\centering
\caption{Component ablation over 30 replicates.}
\label{tab:supp_ablation}
\small
\resizebox{\textwidth}{!}{%
\begin{tabular}{@{}lccccc@{}}
\toprule
Variant & Edges & MCC & FDR & TPR & SHD \\
\midrule
Oracle latent coordinates + DAG & 26.80 & $0.635\;(0.023)$ & 0.261 & 0.598 & 19.27 \\
Observed ALR + DAG & 26.30 & $0.569\;(0.021)$ & 0.320 & 0.537 & 23.70 \\
Recovered coordinates, no final threshold & 26.57 & $0.576\;(0.019)$ & 0.320 & 0.547 & 22.07 \\
Full CoLaDAG & 25.87 & $0.582\;(0.020)$ & 0.305 & 0.544 & 21.77 \\
\bottomrule
\end{tabular}
}
\end{table}

Among implementable variants, full CoLaDAG had the largest mean MCC and the
smallest mean FDR and SHD. The improvements over observed ALR were modest at the
examined sequencing depth.

\subsection{Dimension and graph density}

For the dimension experiment, $d\in\{20,30,50\}$ with edge probability $2/d$.
For the density experiment, $d=30$ and edge probability $c/d$ for
$c\in\{1,2,3\}$. Settings used $n=500,M=10{,}000$ and 20 replicates.

\begin{table}[htbp]
\centering
\caption{Dimension and graph-density robustness.}
\label{tab:supp_grid}
\resizebox{\textwidth}{!}{%
\begin{tabular}{@{}lccc@{}}
\toprule
Setting & CoLaDAG MCC / FDR & Best comparator & Comparator MCC / FDR \\
\midrule
Dimension $d=20$ & 0.597 / 0.306 & PC & 0.424 / 0.518 \\
Dimension $d=30$ & 0.570 / 0.330 & PC & 0.338 / 0.619 \\
Dimension $d=50$ & 0.504 / 0.417 & PC & 0.333 / 0.636 \\
Density $c=1$ & 0.700 / 0.252 & Tabu-Huge & 0.550 / 0.484 \\
Density $c=2$ & 0.573 / 0.341 & PC & 0.360 / 0.594 \\
Density $c=3$ & 0.435 / 0.459 & PC & 0.237 / 0.661 \\
\bottomrule
\end{tabular}
}
\end{table}

CoLaDAG had the largest mean MCC and smallest mean FDR in these six aligned
settings, while performance declined with dimension and density.

\subsection{Final-threshold sensitivity}

The latent recovery and continuous graph output were held fixed within each
replicate while only the final threshold was varied (30 replicates,
$n=300,d=20,M=10{,}000$).

\begin{table}[htbp]
\centering
\caption{Final edge-threshold sensitivity.}
\label{tab:supp_threshold}
\begin{tabular}{lccccc}
\toprule
Threshold & Edges & MCC & FDR & TPR & SHD \\
\midrule
0.00 & 26.23 & 0.573 & 0.319 & 0.541 & 22.07 \\
0.05 & 26.23 & 0.573 & 0.319 & 0.541 & 22.07 \\
0.10 & 26.23 & 0.573 & 0.319 & 0.541 & 22.07 \\
0.20 & 26.23 & 0.573 & 0.319 & 0.541 & 22.07 \\
0.30 & 25.57 & 0.579 & 0.305 & 0.538 & 21.80 \\
0.40 & 23.83 & 0.582 & 0.279 & 0.522 & 21.57 \\
0.50 & 22.10 & 0.591 & 0.243 & 0.509 & 21.13 \\
\bottomrule
\end{tabular}
\end{table}

\begin{figure}[htbp]
\centering
\includegraphics[width=\textwidth]{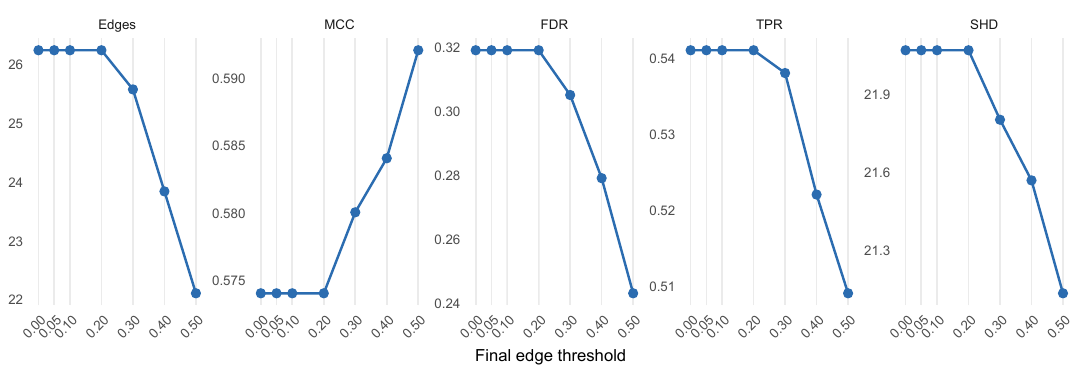}
\caption{Final-threshold sensitivity under the aligned $n=300,d=20$ setting.}
\label{fig:supp_threshold}
\end{figure}

Increasing the threshold reduced mean edge count and FDR while increasing MCC
and reducing SHD, at the cost of lower TPR. This experiment describes a
precision--recall trade-off; it does not select or validate the real-data
threshold $\eta=0.08$.

\section{Real-data Reporting Details}
\label{sec:supp_real}

\subsection{Mouse-block diagnostic budgets}

The primary real-data recovery/refinement budgets were 5/20 and 10/50 DC/ADMM
iterations, respectively. The 20 CoLaDAG block-resampling fits used reduced
budgets of 3/15 and 6/30. The resulting frequencies are therefore approximate
sensitivity diagnostics. Baseline block tuning used 25 resamples for each
configuration and retained all four observations for every selected mouse.

\subsection{Reference-denominator refits}

The reference-sensitivity analysis used the same CoLaDAG parameters and final
$\eta=0.08$ as the primary fit. For each candidate denominator, the raw count
matrix was reordered before the multinomial likelihood was evaluated.
Pseudocount ALR coordinates were constructed for the transformed-data
comparators and for CoLaDAG initialization. Pairwise graph comparisons were
restricted to taxa present as non-reference coordinates in both fits.

\subsection{Reference-node Sensitivity Analysis}
\label{subsec:reference_sensitivity}

A natural concern in ALR-based compositional modeling is that the inferred network
may depend on the arbitrary choice of the denominator taxon. To directly assess
this issue, we conducted a reference-node sensitivity analysis on the real gut
microbiome data. Specifically, we selected the five most abundant retained
compositional parts,
Lachnospiraceae\_NK4A136\_group, Other,
Bacteroides, Alistipes, and Muribaculum, and used
each of them in turn as the ALR reference. For each reference choice, we refit
all directed graph learning methods and compared the resulting graphs pairwise
on the common non-reference taxa. This comparison avoids artificially penalizing
a method simply because the current reference taxon is excluded from the ALR
coordinate system.

Table~\ref{tab:reference_sensitivity} summarizes the resulting pairwise
structural Hamming distances (SHD), edge-overlap rates, and reference-stable
core skeleton counts. Under the aligned tuning protocol, the CoLaDAG graphs
varied substantially in density across references, from 284 directed edges with
Lachnospiraceae\_\allowbreak NK4A136\_\allowbreak group as denominator to 684 directed edges
with Other as denominator. CoLaDAG had mean directed and skeleton
overlap rates of 0.346 and 0.368, respectively; 355 skeleton edges appeared in
at least three of the five reference choices and 92 appeared under all five
choices. These large core counts must be interpreted together with the high
graph densities: they do not imply reference-invariant directed effects. PC and
MMHC showed higher mean skeleton overlap with much sparser graphs. Overall, the
sensitivity analysis confirms that ALR-coordinate graphs are materially
reference-dependent and that skeleton-level summaries are more defensible than
single-reference edge orientations.

\begin{table}[htbp]
    \centering
    \caption{Reference-node sensitivity.}
    \label{tab:reference_sensitivity}
    \small
    \setlength{\tabcolsep}{2pt}
    \begin{tabular}{@{}lrrrrrr@{}}
        \toprule
        Method & \shortstack{Directed\\SHD} & \shortstack{Skeleton\\SHD} &
        \shortstack{Directed\\overlap} & \shortstack{Skeleton\\overlap} &
        \shortstack{Core edges\\$\geq 60\%$} & \shortstack{Core edges\\$=100\%$} \\
        \midrule
        CoLaDAG & 451.0 & 430.8 & 0.346 & 0.368 & 355 & 92 \\
        PC & 51.3 & 26.4 & 0.356 & 0.522 & 38 & 17 \\
        MMHC & 42.6 & 29.6 & 0.349 & 0.497 & 41 & 17 \\
        Tabu-Huge & 96.4 & 70.2 & 0.257 & 0.396 & 66 & 23 \\
        HC-Huge & 91.1 & 70.3 & 0.283 & 0.396 & 67 & 22 \\
        VI-NOTEARS-style & 34.7 & 34.7 & 0.330 & 0.330 & 29 & 7 \\
        \bottomrule
    \end{tabular}
\end{table}

\begin{figure}[htbp]
    \centering
    \includegraphics[width=0.9\textwidth]{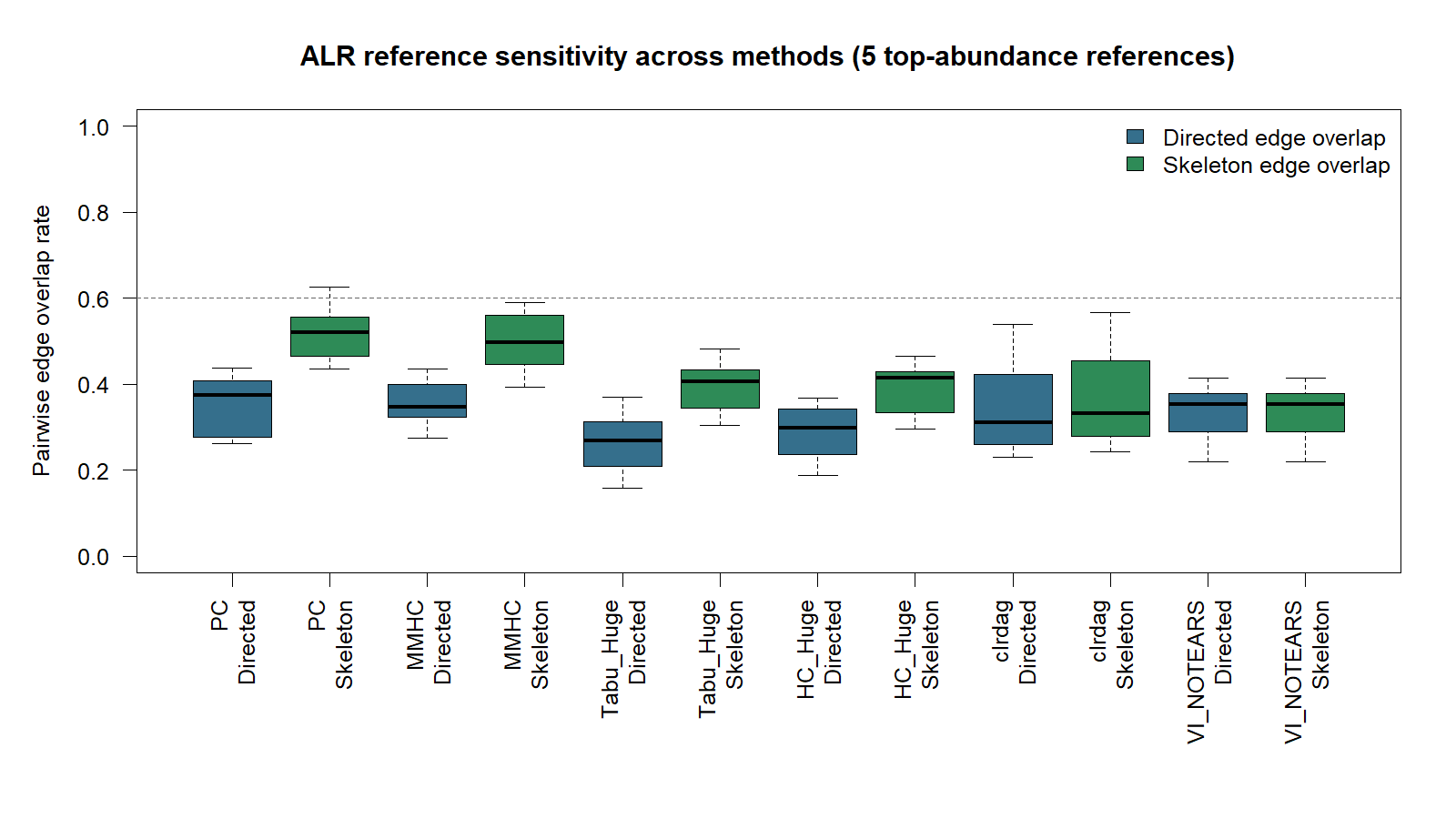}
    \caption{Reference sensitivity across five ALR denominators.}
    \label{fig:reference_sensitivity}
\end{figure}

As presented in Table 3 in the main text, the primary genus-level
analysis produced different graph densities across methods. CoLaDAG selected the
largest candidate edge set in this audited count-likelihood run, whereas PC,
MMHC, VI-NOTEARS-style, HC-Huge, and Tabu-Huge selected sparser candidate graphs under
their specified thresholds. We do not use the full 284-edge CoLaDAG graph as the
main display because such a plot is difficult to interpret. Instead, the
ranked-edge table in Table~4 in the main text, the exposure-group
weight heatmap in Figure~4 in the main text, the retained network
overlay in Figure~5 in the main text, and the reference-sensitivity analysis in
Table~\ref{tab:reference_sensitivity} provide more interpretable views of the
same fitted procedure.

From an ecological perspective, the CoLaDAG genus-level network
highlights several large fitted log-ratio coefficients that can be used to
prioritize follow-up checks. The mouse-block analysis retained only 60 of 284
primary edges at selection frequency 0.60, with 37 additionally meeting the
sign-agreement criterion. The ranked examples are therefore descriptive
coefficient hypotheses rather than stable edge discoveries. The largest
absolute coefficients involved
Intestinimonas, Alistipes, Oscillibacter, Ruminiclostridium group 5,
Rikenellaceae RC9 gut group, and Lachnospiraceae AC2044 group.
Because the ALR denominator is
Lachnospiraceae\_NK4A136\_group, no directed edge incident to that
reference taxon is estimable in the primary graph. These coefficients should be
treated as coordinate-specific prioritization signals, not as evidence of
taxon-level biological mechanisms. The most defensible biological payoff is
therefore a ranked validation plan: quantify the named taxa with targeted assays,
measure candidate fermentation products, and test whether the relations persist
after accounting for mouse and time in a larger experiment.

\subsection{Dose-group slope construction}

The dose displays do not refit group-specific DAGs. The 80 largest absolute
coefficients from the global CoLaDAG fit define a restricted candidate list.
For each candidate and group, the submitted script fits
\[
Z_{\mathrm{child}}=\beta_0+\beta_1Z_{\mathrm{parent}}
+\bm\gamma^\top\mathrm{time}+e.
\]
The reported quantity is $\widehat\beta_1$, a time-adjusted pairwise ALR slope.
It is not a conditional SEM coefficient because the other parents of the child
are absent. Slopes with $|\widehat\beta_1|\geq0.75$ are retained and at most 20
are displayed per group. The conventional intervals below treat 12 observations
as independent and ignore support selection and multiplicity; with only three
mice per group, they are descriptive only.

\begin{landscape}
\begin{scriptsize}
\setlength{\tabcolsep}{1pt}
\setlength\LTleft{0pt}
\setlength\LTright{0pt}
\begin{longtable}{@{}llP{0.19\linewidth}P{0.23\linewidth}P{0.14\linewidth}P{0.16\linewidth}@{}}
\caption{Displayed time-adjusted pairwise ALR slopes by exposure group.}
\label{tab:supp_dose_slopes}\\
\toprule
Group & Rank & From & To & Estimate (standard error) & Naive 95\% interval \\
\midrule
\endfirsthead
\toprule
Group & Rank & From & To & Estimate (standard error) & Naive 95\% interval \\
\midrule
\endhead
\bottomrule
\endfoot
Control & 1 & \path{Intestinimonas} & \path{Dubosiella} & 2.752 (0.780) & [0.908, 4.597] \\
Control & 2 & \path{Intestinimonas} & \path{Rikenellaceae_RC9_gut_group} & 2.123 (1.254) & [-0.842, 5.089] \\
Control & 3 & \path{Intestinimonas} & \path{Ruminococcaceae_UCG-004} & 1.764 (1.030) & [-0.671, 4.198] \\
Control & 4 & \path{Alistipes} & \path{Ruminiclostridium_5} & 1.261 (0.319) & [0.506, 2.015] \\
Control & 5 & \path{Intestinimonas} & \path{Azospirillum_sp._47_25} & 1.207 (1.486) & [-2.308, 4.721] \\
Control & 6 & \path{Sphingomonas} & \path{Rikenellaceae_RC9_gut_group} & 1.184 (0.847) & [-0.819, 3.188] \\
Control & 7 & \path{Oscillibacter} & \path{Angelakisella} & 1.174 (0.996) & [-1.180, 3.528] \\
Control & 8 & \path{Other} & \path{Rikenellaceae_RC9_gut_group} & 1.100 (0.702) & [-0.561, 2.760] \\
Control & 9 & \path{Alistipes} & \path{Bacteroides} & 1.078 (0.287) & [0.401, 1.756] \\
Control & 10 & \path{Desulfovibrio} & \path{Lachnospiraceae_UCG-001} & 1.059 (0.727) & [-0.660, 2.778] \\
Control & 11 & \path{Oscillibacter} & \path{Rikenellaceae_RC9_gut_group} & 1.058 (1.115) & [-1.579, 3.694] \\
Control & 12 & \path{Ruminiclostridium} & \path{Rikenellaceae_RC9_gut_group} & 1.051 (1.103) & [-1.558, 3.660] \\
Control & 13 & \path{Lachnospiraceae_FCS020_group} & \path{Tyzzerella} & 1.050 (0.368) & [0.179, 1.920] \\
Control & 14 & \path{Alistipes} & \path{Dubosiella} & 1.046 (0.378) & [0.153, 1.940] \\
Control & 15 & \path{Desulfovibrio} & \path{[Eubacterium]_xylanophilum_group} & 1.042 (0.637) & [-0.464, 2.547] \\
Control & 16 & \path{Ruminococcaceae_UCG-014} & \path{Dubosiella} & 0.991 (0.178) & [0.570, 1.413] \\
Control & 17 & \path{Lachnoclostridium} & \path{Lachnospira} & 0.980 (0.902) & [-1.152, 3.113] \\
Control & 18 & \path{Ruminiclostridium} & \path{[Eubacterium]_xylanophilum_group} & -0.903 (0.852) & [-2.919, 1.112] \\
Control & 19 & \path{Sphingomonas} & \path{Faecalibacterium} & 0.884 (0.473) & [-0.234, 2.001] \\
Control & 20 & \path{Ruminococcaceae_UCG-014} & \path{Lachnospiraceae_AC2044_group} & 0.855 (0.362) & [-0.001, 1.711] \\
\addlinespace
0.1Ag & 1 & \path{Intestinimonas} & \path{Lachnospiraceae_UCG-001} & -2.075 (0.978) & [-4.388, 0.238] \\
0.1Ag & 2 & \path{Other} & \path{Mucispirillum} & -1.856 (0.909) & [-4.006, 0.293] \\
0.1Ag & 3 & \path{Ruminococcaceae_NK4A214_group} & \path{Lachnospiraceae_AC2044_group} & 1.788 (0.543) & [0.504, 3.072] \\
0.1Ag & 4 & \path{Intestinimonas} & \path{Lachnospiraceae_AC2044_group} & 1.747 (1.250) & [-1.209, 4.703] \\
0.1Ag & 5 & \path{Ruminiclostridium_5} & \path{Parabacteroides} & 1.587 (0.531) & [0.332, 2.842] \\
0.1Ag & 6 & \path{Other} & \path{Rikenellaceae_RC9_gut_group} & 1.569 (0.520) & [0.338, 2.800] \\
0.1Ag & 7 & \path{Alistipes} & \path{Lachnospiraceae_AC2044_group} & 1.550 (1.073) & [-0.988, 4.089] \\
0.1Ag & 8 & \path{Ruminiclostridium_5} & \path{Lachnospiraceae_AC2044_group} & 1.538 (1.367) & [-1.693, 4.770] \\
0.1Ag & 9 & \path{Lachnoclostridium} & \path{Lachnospira} & 1.506 (0.556) & [0.191, 2.820] \\
0.1Ag & 10 & \path{Intestinimonas} & \path{Azospirillum_sp._47_25} & 1.484 (0.752) & [-0.295, 3.263] \\
0.1Ag & 11 & \path{Ruminococcaceae_UCG-014} & \path{Lachnospiraceae_AC2044_group} & 1.476 (0.792) & [-0.398, 3.349] \\
0.1Ag & 12 & \path{Other} & \path{Helicobacter} & 1.474 (1.116) & [-1.166, 4.114] \\
0.1Ag & 13 & \path{Lachnospiraceae_FCS020_group} & \path{Lachnospiraceae_UCG-001} & -1.464 (0.630) & [-2.953, 0.025] \\
0.1Ag & 14 & \path{Intestinimonas} & \path{Ruminococcaceae_UCG-004} & 1.436 (0.610) & [-0.007, 2.879] \\
0.1Ag & 15 & \path{Ruminococcaceae_NK4A214_group} & \path{Lachnospiraceae_UCG-001} & -1.429 (0.547) & [-2.723, -0.135] \\
0.1Ag & 16 & \path{Intestinimonas} & \path{Dubosiella} & -1.350 (1.014) & [-3.747, 1.047] \\
0.1Ag & 17 & \path{Bacteroides} & \path{Alloprevotella} & 1.331 (0.407) & [0.369, 2.294] \\
0.1Ag & 18 & \path{Butyricicoccus} & \path{Dubosiella} & -1.284 (0.283) & [-1.954, -0.614] \\
0.1Ag & 19 & \path{Oscillibacter} & \path{Mucispirillum} & 1.238 (0.724) & [-0.474, 2.950] \\
0.1Ag & 20 & \path{Lachnoclostridium} & \path{Lachnospiraceae_UCG-001} & -1.215 (1.060) & [-3.721, 1.292] \\
\addlinespace
2Ag & 1 & \path{Other} & \path{Rikenellaceae_RC9_gut_group} & 4.207 (1.752) & [0.065, 8.349] \\
2Ag & 2 & \path{Oscillibacter} & \path{Mucispirillum} & 3.010 (1.257) & [0.038, 5.982] \\
2Ag & 3 & \path{Alistipes} & \path{Mucispirillum} & 2.970 (0.669) & [1.389, 4.551] \\
2Ag & 4 & \path{Intestinimonas} & \path{Lachnospiraceae_AC2044_group} & 2.708 (1.324) & [-0.424, 5.839] \\
2Ag & 5 & \path{Intestinimonas} & \path{Azospirillum_sp._47_25} & 2.554 (1.663) & [-1.379, 6.486] \\
2Ag & 6 & \path{Lachnoclostridium} & \path{Marvinbryantia} & 2.322 (1.312) & [-0.780, 5.423] \\
2Ag & 7 & \path{Sphingomonas} & \path{Rikenellaceae_RC9_gut_group} & 1.980 (0.570) & [0.632, 3.328] \\
2Ag & 8 & \path{Sphingomonas} & \path{ASF356} & 1.946 (0.442) & [0.902, 2.991] \\
2Ag & 9 & \path{Ruminiclostridium} & \path{[Eubacterium]_xylanophilum_group} & -1.626 (0.434) & [-2.653, -0.599] \\
2Ag & 10 & \path{Oscillibacter} & \path{[Eubacterium]_xylanophilum_group} & -1.567 (0.729) & [-3.291, 0.157] \\
2Ag & 11 & \path{Faecalibacterium} & \path{[Ruminococcus]_gnavus_group} & 1.562 (0.909) & [-0.587, 3.712] \\
2Ag & 12 & \path{Ruminococcaceae_UCG-010} & \path{Parabacteroides} & -1.456 (0.466) & [-2.557, -0.355] \\
2Ag & 13 & \path{Oscillibacter} & \path{Butyricicoccus} & 1.277 (0.489) & [0.121, 2.433] \\
2Ag & 14 & \path{Oscillibacter} & \path{Rikenellaceae_RC9_gut_group} & 1.247 (1.413) & [-2.094, 4.588] \\
2Ag & 15 & \path{Bacteroides} & \path{Alloprevotella} & 1.232 (0.554) & [-0.079, 2.543] \\
2Ag & 16 & \path{Oscillibacter} & \path{Bilophila} & 1.224 (0.735) & [-0.514, 2.962] \\
2Ag & 17 & \path{Lachnospiraceae_FCS020_group} & \path{Tyzzerella} & 1.217 (0.689) & [-0.411, 2.846] \\
2Ag & 18 & \path{Lachnoclostridium} & \path{Lachnospira} & 1.216 (1.433) & [-2.172, 4.603] \\
2Ag & 19 & \path{Ruminococcaceae_UCG-014} & \path{Lachnospiraceae_AC2044_group} & -1.122 (0.737) & [-2.864, 0.620] \\
2Ag & 20 & \path{Roseburia} & \path{[Eubacterium]_xylanophilum_group} & -1.049 (0.615) & [-2.504, 0.406] \\
\addlinespace
40Ag & 1 & \path{Lachnoclostridium} & \path{A2} & 2.092 (1.154) & [-0.636, 4.820] \\
40Ag & 2 & \path{Intestinimonas} & \path{Ruminococcus_1} & 1.973 (0.534) & [0.709, 3.237] \\
40Ag & 3 & \path{Intestinimonas} & \path{Butyricicoccus} & 1.921 (0.434) & [0.894, 2.948] \\
40Ag & 4 & \path{Lachnoclostridium} & \path{Lachnospira} & 1.891 (0.898) & [-0.232, 4.014] \\
40Ag & 5 & \path{Lachnoclostridium} & \path{Lachnospiraceae_UCG-001} & -1.794 (1.432) & [-5.180, 1.592] \\
40Ag & 6 & \path{Bilophila} & \path{Dubosiella} & -1.657 (0.518) & [-2.881, -0.433] \\
40Ag & 7 & \path{Lachnoclostridium} & \path{Marvinbryantia} & 1.588 (1.612) & [-2.223, 5.399] \\
40Ag & 8 & \path{Sphingomonas} & \path{Rikenellaceae_RC9_gut_group} & 1.374 (0.436) & [0.344, 2.405] \\
40Ag & 9 & \path{Intestinimonas} & \path{Oscillibacter} & 1.368 (0.202) & [0.890, 1.846] \\
40Ag & 10 & \path{Oscillibacter} & \path{Butyricicoccus} & 1.295 (0.304) & [0.577, 2.013] \\
40Ag & 11 & \path{Enterorhabdus} & \path{Prevotellaceae_UCG-001} & 1.280 (0.137) & [0.956, 1.605] \\
40Ag & 12 & \path{Roseburia} & \path{Faecalibaculum} & 1.244 (0.754) & [-0.538, 3.027] \\
40Ag & 13 & \path{Intestinimonas} & \path{Ruminococcaceae_UCG-004} & 1.228 (0.586) & [-0.158, 2.615] \\
40Ag & 14 & \path{Intestinimonas} & \path{Dubosiella} & -1.132 (0.858) & [-3.161, 0.898] \\
40Ag & 15 & \path{Other} & \path{Rikenellaceae_RC9_gut_group} & 1.111 (0.534) & [-0.152, 2.374] \\
40Ag & 16 & \path{Sphingomonas} & \path{Faecalibacterium} & 1.092 (0.195) & [0.631, 1.553] \\
40Ag & 17 & \path{Lachnospiraceae_FCS020_group} & \path{Lachnospiraceae_UCG-001} & 1.081 (0.783) & [-0.771, 2.934] \\
40Ag & 18 & \path{Lachnoclostridium} & \path{Enterorhabdus} & 1.034 (0.657) & [-0.520, 2.588] \\
40Ag & 19 & \path{Intestinimonas} & \path{Lachnospiraceae_AC2044_group} & 1.029 (1.900) & [-3.465, 5.523] \\
40Ag & 20 & \path{Ruminiclostridium_5} & \path{Ruminococcus_1} & 1.023 (0.336) & [0.230, 1.817] \\

\end{longtable}
\end{scriptsize}

The pooled ``Other'' feature appears in the numerical list because it was part
of the analyzed composition, but it is not interpreted as a biological taxon.

\end{landscape}

\end{document}